\documentclass[submission,Phys]{SciPost}

\usepackage[utf8]{inputenc} % input encodings (allow UTF-8 input)
\usepackage[T1]{fontenc} 	% selecting font encodings (use 8-bit T1 fonts)
\usepackage[english]{babel} % multilingual support (English language/hyphenation)

\usepackage[bitstream-charter]{mathdesign}
\usepackage{geometry} 		% flexible and complete interface to document dimensions
\usepackage{amsmath} 		% math package (American Mathematical Society)
\usepackage{mathtools} 		% math package (fixes various deficiencies of amsmath)
\usepackage{float} 			% floating objects such as figures and tables
\usepackage{graphicx} 		% enhanced support for graphics
\usepackage{tabularx} 		% tabulars with adjustable-width columns
\usepackage{booktabs} 		% professional-quality tables
\usepackage{color, xcolor} 	% foreground and background colour management
\usepackage{pdfpages} 		% inclusion of external multi-page PDF documents
\usepackage{extarrows} 		% extra arrows beyond those provided in amsmath
\usepackage{multirow} 		% create tabular cells spanning multiple rows
\usepackage{multicol} 		% intermix single and multiple columns
\usepackage[subrefformat=parens]{subcaption}
\usepackage{enumitem} 		% control layout of itemize, enumerate, description
\usepackage{xspace} 		% define commands that appear not to eat spaces
\usepackage{stackrel} 		% enhancement to the \stackrel command
\usepackage{tikz} 			% create PostScript and PDF graphics
\usetikzlibrary{calc}
\usepackage{braket} 		% Dirac bra-ket notation
\usepackage{bm} 			% access bold symbols in maths mode
\usepackage{tensor} 		% typeset tensors (tensor-style super- and subscripts)
\usepackage{slashed} 		% slash through characters (Feynman slash notation)
\usepackage{siunitx} % SI units package (typesetting values with units)
\usepackage{lastpage} 		% reference last page
\usepackage{cite} 			% improved citation handling
\usepackage[normalem]{ulem} % package for underlining
\usepackage{fontawesome} 	% access to web-related icons
\usepackage{tocloft} 		% control over the typography of the TOC, LOF, and LOT
\usepackage{titlesec} 		% interface to sectioning commands (various title styles)
\usepackage{doi} 			% create correct hyperlinks for DOI numbers
\usepackage{hyperref} 		% hypertext links (handel cross-referencing commands)
\usepackage[most]{tcolorbox} 					% coloured and framed text boxes
\usepackage[nameinlink, capitalize]{cleveref} 	% intelligent cross-referencing
\usepackage[nottoc, notlot, notlof]{tocbibind} 	% add references/index/contents to TOC
\usepackage[ruled, vlined]{algorithm2e} 		% floating algorithm environment
\usepackage{makecell}
\usepackage{setspace}
\usepackage{orcidlink}

\makeatletter
\def\BState{\State\hskip-\ALG@thistlm}
\makeatother

\makeatletter
\@ifundefined{pdfoutput}{}{\DeclareGraphicsRule{*}{mps}{*}{}}
\makeatother

\makeatletter
\DeclareRobustCommand*{\bfseries}{%
   \not@math@alphabet\bfseries\mathbf
   \fontseries\bfdefault\selectfont
   \boldmath
}
\makeatother

\hypersetup{
	colorlinks=true, 			% false: boxed links, true: colored links
	linkcolor={red!50!black}, 	% color of internal links (sections, pages, etc.)
	citecolor={blue!50!black}, 	% color of citation links (links to bibliography)
	urlcolor={blue!80!black} 	% color of URL links (external links)
} 
\DeclareSymbolFont{usualmathcal}{OMS}{cmsy}{m}{n}
\DeclareSymbolFontAlphabet{\mathcal}{usualmathcal}

\SetArgSty{textnormal}
\SetKwComment{Comment}{{\small\#}~}{}
\SetCommentSty{mycommfont}

\setitemize{itemsep=0pt, parsep=0pt} 				% adjust itemize environment
\setenumerate{itemsep=0pt, parsep=0pt} 				% adjust enumerate environment
\renewcommand{\thefootnote}{\fnsymbol{footnote}} 	% footnote symbol
\setitemize{itemsep=2pt,topsep=2pt,parsep=0pt,partopsep=0pt,leftmargin=*}
\setenumerate{itemsep=0pt,topsep=2pt,parsep=0pt,partopsep=0pt,labelindent=3pt,leftmargin=*}
\usepackage{placeins}

\newcommand{\ltsima}{$\; \buildrel < \over \sim \;$}
\newcommand{\lsim}{\lower.5ex\hbox{\ltsima}}
\newcommand{\gtsima}{$\; \buildrel > \over \sim \;$}
\newcommand{\gsim}{\lower.5ex\hbox{\gtsima}}

\newcommand{\dd}{\mathrm{d}}

\newcommand{\loss}{\mathcal{L}} 	% loss value
\newcommand{\arXiv}[2][]{%
	\ifthenelse{\equal{#1}{}}%
	{\href{http://arxiv.org/abs/#2}{arXiv:#2}}%
	{\href{http://arxiv.org/abs/#2}{arXiv:#2~[#1]}}}

\def\slashchar#1{\setbox0=\hbox{$#1$}           % set a box for #1
   \dimen0=\wd0                                 % and get its size
   \setbox1=\hbox{/} \dimen1=\wd1               % get size of /
   \ifdim\dimen0>\dimen1                        % #1 is bigger
      \rlap{\hbox to \dimen0{\hfil/\hfil}}      % so center / in box
      #1                                        % and print #1
   \else                                        % / is bigger
      \rlap{\hbox to \dimen1{\hfil$#1$\hfil}}   % so center #1
      /                                         % and print /
   \fi}

\newcommand{\tikznode}[2]{%
\ifmmode%
\tikz[remember picture,baseline=(#1.base),inner sep=0pt] \node (#1) {$#2$};%
\else
\tikz[remember picture,baseline=(#1.base),inner sep=0pt] \node (#1) {#2};%
\fi}

\def\mathswitchr#1{\relax\ifmmode{\text{#1}}\else$\text{#1}$\xspace\fi}
\def\mathswitch#1{\relax\ifmmode#1\else$#1$\xspace\fi}

\graphicspath{{./figures/}}
\begin{document}

\begin{center}{\Large \textbf{
      PINNferring the Hubble Function with 
      Uncertainties
}}\end{center}

\begin{center}
Lennart R\"over \orcidlink{0000-0002-4248-8329}\textsuperscript{1},
Bj\"orn Malte Sch\"afer \orcidlink{0000-0002-9453-5772}\textsuperscript{1,3}, and
Tilman Plehn\orcidlink{0000-0001-5660-7790}\textsuperscript{2,3}
\end{center}

\begin{center}
{\bf 1} Zentrum f\"ur Astronomie der Universit\"at Heidelberg, Astronomisches Rechen-Institut, Heidelberg, Germany \\
{\bf 2} Institut f\"ur Theoretische Physik, Universit\"at Heidelberg, Germany\\
{\bf 3} Interdisciplinary Center for Scientific Computing (IWR), Universit\"at Heidelberg, Germany
\end{center}

%\begin{center}
%\today
%\end{center}

\section*{Abstract}
{\bf The Hubble function characterizes a given Friedmann-Robertson-Walker spacetime and can  be related to the densities of
  the cosmological fluids and their equations of state. We show how physics-informed neural networks (PINNs)
  emulate this dynamical system and provide fast predictions of the luminosity distance for a given choice of densities and equations of state, as needed for the analysis of supernova data. We use this emulator to perform a model-independent and parameter-free reconstruction of the Hubble function on the basis of supernova data. As part of this study, we develop and validate an uncertainty treatment for PINNs using a heteroscedastic loss and repulsive ensembles.}

% For convenience during refereeing: line numbers
%\linenumbers

\vspace{10pt}
\noindent\rule{\textwidth}{1pt}
\tableofcontents\thispagestyle{fancy}
\noindent\rule{\textwidth}{1pt}
\vspace{10pt}

\clearpage
%%%%%%%%%%%%%%%%%%%%%%%%%%%%%%%%%%%%%%%%%%%%%%%%
\section{PINNtroduction}
\label{sec:intro}

Friedmann-Robertson-Walker spacetimes are entirely characterized by
their Hubble function $H(a) = \dot{a}/a$ as a consequence of the
cosmological principle, which requires homogeneity and isotropy on
cosmological scales. Substitution of the Friedmann-Robertson-Walker
line element into the gravitational field equation allows to relate
the Hubble function to the densities of the cosmological fluids and
their respective equations of state. Probing the Hubble function not
only addresses these densities and their equations of state, but also
interactions between the fluids and their possible non-adiabatic
evolution. The Hubble function itself only relies on symmetry
assumptions and is in fact not restricted by theory to follow any
a-priori parameterization. As such, it is  possible to
reconstruct it without recourse to a specific cosmological model,
like assumptions about the gravitational field equation or
specific properties of the cosmological fluids.

Perhaps the most direct probe of cosmic evolution out to redshifts
beyond unity are supernovae of type Ia~\cite{linder_cosmic_2005,
  barnes_influence_2005}. They allow constraints on the evolution of
luminosity distance with redshift, and therefore indirectly
 on the Hubble function, from which the luminosity distance
follows after an integration. A typical effect indicative of repulsive
gravity on large scales are systematically darker supernovae as they
approach the cosmic horizon. These effects are associated with
cosmological fluids with equations of state $w < -1/3$, typical for
dark energy or the cosmological constant. In many cases, the
equation of state is an immutable property of the cosmic fluid, for
instance $w=0$ for matter, and often one works with constant or
linearly evolving equations of state for dark
energy~\cite{chevallier_accelerating_2001, linder_exploring_2003},
even though there are compelling arguments for evolving dark
energy~\cite{WetterichQuintessence, PeeblesQuintessence,
  linder_dynamics_2008, tsujikawa_quintessence:_2013,
  mortonson_dark_2013}.

Our approach relies only on the
symmetry principles for spacetime and derives the Hubble function
$H(a)$ free of any parameterization directly from data. 
We achieve this with physics-informed neural
networks (PINNs)~\cite{DBLP:journals/corr/abs-1711-10561,Piscopo:2019txs,Araz:2021hpx,DBLP:journals/corr/abs-2111-03794,
  DBLP:journals/corr/abs-2201-05624, Hao2022PhysicsInformedML}, similar techniques have been used in Refs.~\cite{Anderson_2021, Douglas:2020hpv, Jejjala_2022, Larfors_2022, Berglund:2022gvm}.
They absorb the space of solutions of a
differential equation with a free representation of the Hubble
function given by a second neural network. This is necessary
because the Hubble function is not directly observable. The
relevant observable is the
luminosity distance, a weighted integral over the
inverse Hubble function. The PINN needs to learn a
fast prediction of the luminosity distance for a given Hubble function,
which is represented by another neural network. 

In physics, a number without an error bar cannot describe a measurement 
and it also cannot describe a prediction. This means, we need to 
develop an uncertainty estimate for the learned function encoded in the PINNs.
Sources of uncertainties of a trained neural network include 
statistical limitation of the training sample, stochasticity or noise 
of the training data, theory uncertainties in simulated training data, or
a lack of flexibility of the network architecture~\cite{Plehn:2022ftl}. We employ 
two different methods to learn an error band on the network prediction, 
a heteroscedastic loss function~\cite{LeHeteroscedastic, Gal2016UncertaintyID} 
and repulsive ensembles~\cite{DAngelo2021RepulsiveDE}. 

Our paper starts with a brief introduction to PINNs and the two 
methods to also learn an uncertainty in Sec.~\ref{sec:pinn}.
To the best of our knowledge, we apply repulsive ensembles to 
a cosmological problem for the first time, so we include a more detailed 
derivation in Sec.~\ref{sec:pinn_repuls}. 
In Sec.~\ref{sect:PINNasEmulator} we train a PINN emulator and study 
its behavior. Finally, we show how the combination 
of two networks can be used to extract the Hubble function from the 
luminosity in Sec.~\ref{sect:PINNference}. We find that our inference works 
great, up to redshifts where the experimental uncertainties do not allow 
for a meaningful extraction anymore. %Bummer.

%%%%%%%%%%%%%%%%%%%%%%%%%%%%%%%%%%%%%%%%%%%%%%%%
\section{PINNcertainties}
\label{sec:pinn}

The idea behind physics-informed neural
networks~\cite{DBLP:journals/corr/abs-1711-10561,Piscopo:2019txs,Araz:2021hpx,
  DBLP:journals/corr/abs-2111-03794,
  DBLP:journals/corr/abs-2201-05624, Hao2022PhysicsInformedML} is to understand and reproduce training data more efficiently by learning it as a 
  solution to a
differential equation. For an ordinary differential equation,
\begin{align}
  \dot{u}(t) = F(u,t)
  \qquad \text{with initial conditions} \qquad
  u(t=0) = u_0 \; ,
  \label{eq:generalODE}
\end{align}
their MSE loss consists of two terms,
\begin{align}
  \loss  = (1-\beta ) &\loss_\text{IC}  + \beta \loss_\text{ODE} \notag \\
  \text{with} \qquad 
  \loss_\text{IC}
  &= \left[ u_\theta(t=0) - u_0 \right]^2 \notag \\
  %= \left|\mathbf{u}^i_{\theta} - \mathbf{u}^i_\text{true}\right|^2
  \loss_\text{ODE}
  &= \left[ \dot{u}_{\theta}(t) - F(u_\theta,t)\right] ^2 \; .
  %= \left|\dot{\mathbf{u}}_{\theta}(t) - \dot{\mathbf{u}}_\text{true}(t)\right|^2.
\label{eq:generalLoss}
\end{align}
PINNs form, together with neural differential equations and neural operators, a group of machine learning methods relating neural networks to solutions of differential equations. Here, PINNs learn a prediction for a given parameter choice without really solving an ODE at the stage of evaluation. Neural ODEs \cite{2018arXiv180607366C} use neural networks as part of a system of differential equations that is solved with conventional methods. Neural operators \cite{2018arXiv181008552P} provide a parameterized mapping of e.g. initial conditions to a state at a given time, but can be used in a more general context.

The first term drives the PINN to fulfill the initial conditions, and
can be used without any additional training data. The second term
ensures that the network approximates a solution to the differential
equation.  The parameter $\beta$ balances the two contributions.

The PINN training through the ODE loss uses two kinds of data. First, 
unlabeled or residual data points consist of points in time, 
where the differential equation is evaluated during the
training~\cite{Raissi2017PhysicsID}. For the ODE loss these time 
points determine where the differential equation is evaluated. Second,
labeled time points can include other information, in our case 
the corresponding true values for $u(t)$ and $\dot{u}(t)$.

%%%%%%%%%%%%%%%%%%%%%%%%%%%%%%%%%%%%%%%%%%%%%%%%
\subsection{Toy example}
\label{sec:pinn_harmonic}

We demonstrate some properties of PINNs for a simple harmonic
oscillation. This includes the quality of the approximation for an
increasing number of residual points, the effect of including labeled
data, and how to estimate uncertainties using a heteroscedastic loss
and repulsive ensembles.  Our toy model is defined by the differential
equation in two dimensions,
\begin{align}
  \ddot{u} + \frac{u}{2} = 0
  \qquad \text{with} \qquad
  u(0) = \begin{pmatrix} 1 \\ 0 \end{pmatrix}
  \qquad
  \dot{u}(0) = \begin{pmatrix} 0 \\ 1 \end{pmatrix} \; .
  \label{eq:simpleHarmonicOsci}
\end{align}
The PINNs are trained on a first-order ODE describing the evolution of
the vector $(u,\dot{u})$, where the two dimensions are independent
of each other. This has the advantage of slightly faster training,
but it sacrifices the guaranteed relation between $u$ and $\dot{u}$.
For all results, we show one of the two components $u_{1,2}(t)$.%, where 
%$u_2(t) \equiv \dot{u}(t)$.

As a complication, our harmonic oscillator,
Eq.\eqref{eq:simpleHarmonicOsci}, has a trivial solution $u(t) =
0$. For this solution $\loss_\text{IC}$ is not minimal, but
$\loss_\text{ODE}$ does not lead to any gradient. Only the coupled
training with both loss terms allows us to construct a non-trivial 
solution, albeit 
including some kind of oscillation with a decreasing
amplitude over time.

%%%%%%%%%%%%%%%%%%%%%%%%%%%%%%%%%%%%%%%%%%%%%%%%
\subsubsection*{Unlabeled or residual data}

\begin{figure}[t]
  \includegraphics[width = 0.495\textwidth]{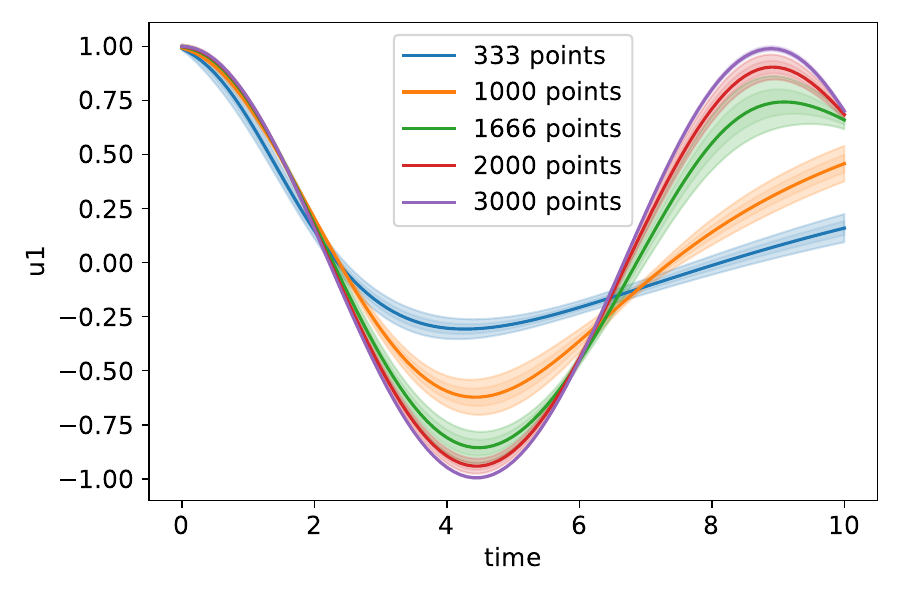}
  \includegraphics[width = 0.495\textwidth]{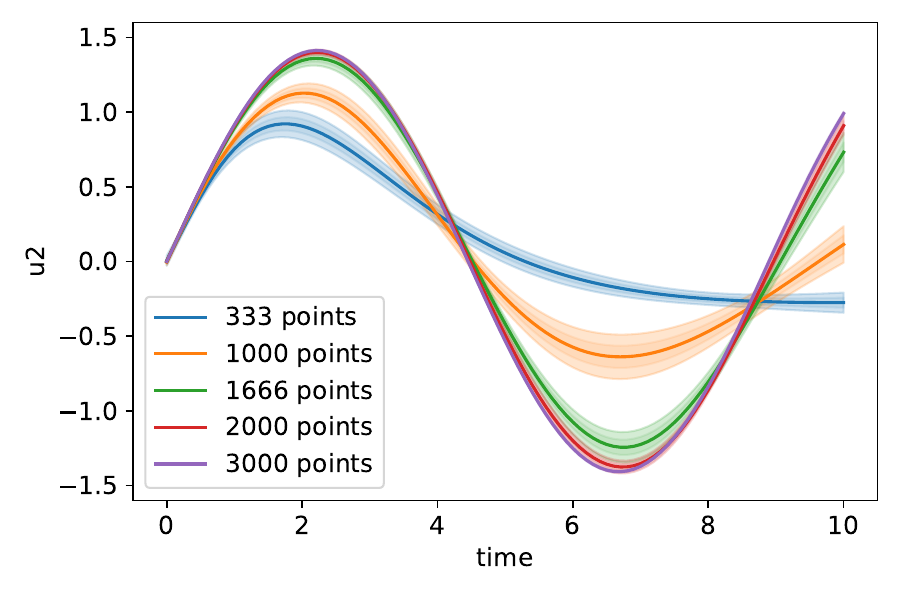}
  \caption{Learned harmonic oscillator, $u(t)$ on the left and $\dot{u}(t)$ on the right, for
    a varying number of uniformly distributed residual points. For the
    ensemble spread we train 10 independent models trained on
    different data points.}
 \label{fig:harmonicReducedComplexity}
\end{figure}

As a first test, we look at the effect of the number of residual
points and introduce ensembles for uncertainty estimation. 
Our basic architecture consists of five
layers with 200 nodes per hidden layer. 
All our networks are written in PyTorch~\cite{Paszke2019PyTorchAI}. 
The training uses the ADAM
optimizer~\cite{Kingma2014AdamAM} in a batch learning setup. For the
loss, we choose equal contributions, $\beta=1/2$. We train
ten networks on 333, 1000, 1666, 2000, and 3000 uniformly distributed
residual points in $t \in [0,10]$. The means and standard deviations
of this ensemble are shown in
Fig.~\ref{fig:harmonicReducedComplexity}. 

%-----------------------------------------------
%-----------------------------------------------

As expected, the approximation improves when the number of residual
points increases. The initial condition is learned even from a small numbers of
residual points, but a good prediction at later times requires more
training points for the ODE loss. The reason is that for later times the
network has to describe a time range rather than just a fixed vector.
The trivial solution $u(t) = 0$, typically close to the network initialization,
give the network the option to learn a shape which approximates the trivial
solution with a decreasing amplitude at late times.  For more residual
points, the agreement with the true solution improves quantitatively
at early times and qualitatively at late times. 

We also see that the uncertainty from the network ensemble does not
capture the poor agreement with the true solution.  The different
networks appear to be drawn to the same local minimum in the loss
function even for different set of residual points.

%%%%%%%%%%%%%%%%%%%%%%%%%%%%%%%%%%%%%%%%%%%%%%%%
\subsubsection*{Labeled data}

%-----------------------------------------------
\begin{figure}[b!]
    \includegraphics[width = 0.495\textwidth]{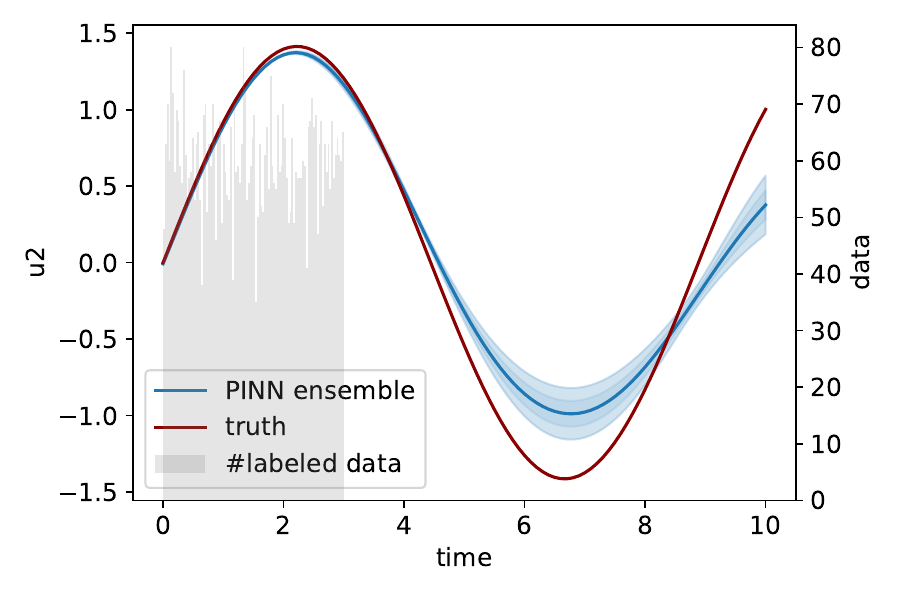}
    \includegraphics[width = 0.495\textwidth]{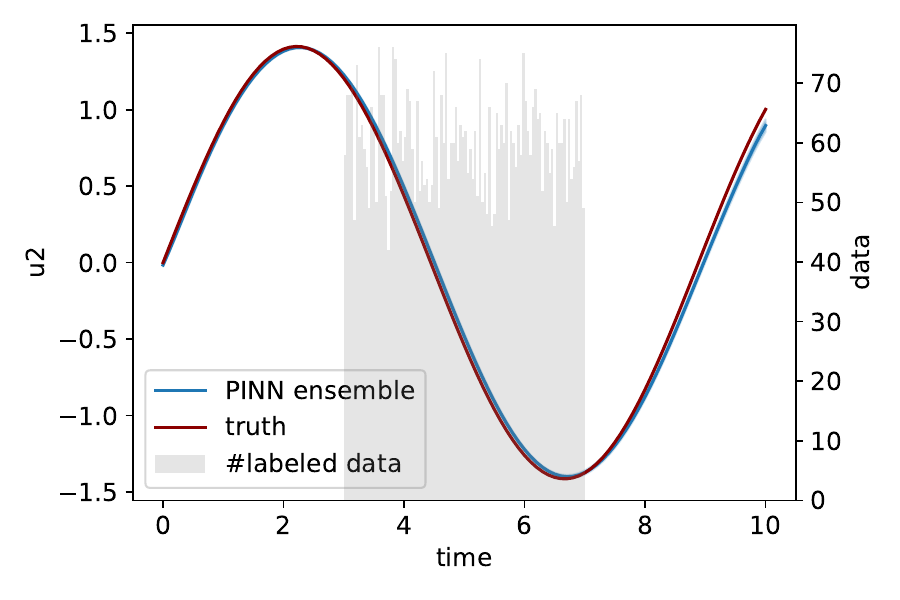}
    \caption{Learned harmonic oscillator adding labeled data
      points at small times (left) and intermediate times (right). 
      The light histogram gives the distribution of training 
      points.  For the ensemble spread we train 10 
      independent models on different
      residual data points.}
 \label{fig:harmonicLeftRightData}
\end{figure} 
%-----------------------------------------------

For many physics problems, the training data can include more information
than just a set of points in time. To judge the impact of this 
data in this section, 
we start with 1000 residual points $t_i$ and combine them
with 6000 labeled points $(t,u,\dot u)_i$.
This additional information can then be used in the ODE loss of 
Eq.\eqref{eq:generalLoss} directly. 
We could 
combine the two kinds of training data by pre-training the usual
network using the labeled
data points. Instead, we train the network alternatingly, where one 
step minimizes
the MSE between the network prediction and the labels of the labeled
data, and a second step minimizes the PINN loss from
Eq.\eqref{eq:generalLoss} on the residual points. We can consider 
training on labeled data as standard network training
samples on existing data samples, while unlabeled or residual
point first generate the information for the network training, 
in analogy to online training. In particle physics, efficient 
integration and sampling builds on a very similar combination of 
online  and buffered or sample-based 
training~\cite{Heimel:2022wyj,Heimel:2023ngj}.

For the harmonic oscillator and its trivial solution it is clear that
uniformly distributed labeled data points are not
optimal.  In Fig.~\ref{fig:harmonicLeftRightData} we show how the PINN
training improves when we include labeled data in specific time
windows, while the unlabeled data remains distributed uniformly.

The left panel shows that
6000 labeled data points close to the initial condition yields a
significant improvement in the region of the labeled data
points. Additionally, there is a small time interval where the PINNs
learn a sensible extrapolation, breaking down at later times. 
The ensemble uncertainties do not cover any of the deviations
from the true solution. In the right panel we position the labeled
points at later times.  Combined with the IC-loss this allows the
networks to learn a good approximation over the entire time range.  If
we consider the initial condition as labeled data as well, this setup
reduces our problem to a simple interpolation. Because the gap between the
initial condition and the additional labeled points does not cover
the first maximum of the oscillation, its position is captured by
the PINN loss.

%%%%%%%%%%%%%%%%%%%%%%%%%%%%%%%%%%%%%%%%%%%%%%%%
\subsection{Gaussian likelihood with errors}
\label{sec:pinn_alea}

The problem with the MSE loss in Eq.\eqref{eq:generalLoss} is that 
it should be related to 
the probability of the network weights to reproduce
the training data, $p(\theta|x_\text{train})$.\footnote{For all ML-related arguments we follow the conventions of the Heidelberg lecture notes, Ref.~\cite{Plehn:2022ftl}.} We usually do not
have access to this probability, but we can use Bayes' theorem to replace
it with a likelihood and a prior, ignoring
the $\theta$-independent evidence~\cite{Plehn:2022ftl}. We will show
this derivation in more detail in Sec.~\ref{sec:pinn_repuls} and
assume for now that a loss should be given by the likelihood,
\begin{align}
  \loss \sim -\log p(x_\text{train}|\theta) \; .
\end{align}
When we compute such a likelihood, physics observations and theory
predictions come with uncertainties. Just like for a fit, these
uncertainties should be part of the network training. The way to
add them to the MSE loss is by noting that the MSE is the
negative logarithm of a Gaussian likelihood assuming a constant
uncertainty.

%%%%%%%%%%%%%%%%%%%%%%%%%%%%%%%%%%%%%%%%%%%%%%%%
\subsubsection*{Heteroscedastic loss} 

This leads us directly to uncertainty quantification for
PINNs via a Gaussian
likelihood with a learned uncertainty function $\sigma_\theta$. 
For a simple, one-dimensional problem this means
\begin{align}
  \loss_\text{het}
  &= - \log \left[ \frac{1}{\sigma_\theta(x)} \;
    \exp \left( - \frac{|f_\theta(x) - f(x)|^2}{2 \sigma_\theta(x)^2} \right) \right]
  + \cdots
  \notag \\
  &= \frac{|f_\theta(x) - f(x)|^2}{2 \sigma_\theta(x)^2} 
  + \log \sigma_\theta(x) + \cdots \;
  \label{eq:hetero_first}
\end{align}
During training, the network can either decrease the numerator of the MSE term or
increase the denominator at the expense of increasing the
normalization terms, so it learns $\sigma_\theta$ by
balancing two explicit loss terms pushing the uncertainty estimate into opposite directions. 
This loss is referred as the
heteroscedastic loss~\cite{LeHeteroscedastic, Gal2016UncertaintyID}.  The
dots include the relevant prior and the irrelevant evidence and
normalization constants from Bayes' theorem and the Gaussian
likelihood.

Unlike Bayesian networks, the heteroscedastic loss does not
distinguish between different sources of uncertainties, for instance
statistical limitations and stochasticity of the
training data, or a limited expressivity of the networks~\cite{Gal2016UncertaintyID,Bollweg:2019skg,Kasieczka:2020vlh,Plehn:2022ftl}.  Still, we will see that it 
is well suited to describe
uncertainties from statistically limited or noisy data, in analogy to a
fit maximizing a Gaussian likelihood. 
%Because the heteroscedastic loss
%typically pushes the envelope of networks in a given minimum of the
%loss landscape, we expect the family of functions it describes to be
%fairly narrow. 
An open question is if it gives a reliable uncertainty estimate in 
the case of insufficient network architectures. On the one hand the
training will focus on a, possibly, local minimum 
in the loss landscape, on the other hand the loss does encourage 
large values of $\sigma$ in case the MSE-like numerator of the 
log-likelihood loss cannot be further reduced.

For the PINN problem of Eq.\eqref{eq:generalODE}, approximating a $d$-dimensional function based on
$N$ residual data points, the heteroscedastic loss gives the
likelihood for the network parameters to describe a solution to the
differential equation, 
\begin{align}
  \loss_\text{IC,het}
  &= \frac{1}{N} \sum_{i = 1}^N \sum_{k = 1}^d 
    \left[ \frac{ | u_{\theta,k}(t_i=0)- u_{0,k} |^2}
         {2 \sigma_{\theta,k}(t_i=0)^2}
         + \log\sigma_{\theta,k}(t_i=0) \right] \notag \\
  \loss_\text{ODE,het}
  &= \frac{1}{N} \sum_{i = 1}^N \sum_{k = 1}^d
    \left[ \frac{ |\dot{u}_{\theta,k}(t_i)-F_k (u_\theta(t_i)) |^2}
         {2 \sigma_{\theta,k}(t_i)^2}
         + \log\sigma_{\theta,k}(t_i) \right] \; .
  \label{eq:heteroscedasticPINN}
\end{align}
In this form, the widths $\sigma_{\theta,k}$ describe how constraining the residual points
are.

In complete analogy to residual points, we can implement a
heteroscedastic loss for the labeled points. In that case we 
start from the same regression loss as for the initial condition
and introduce a learned uncertainty.
This heteroscedastic uncertainty is implemented by
doubling the number of output parameters of the network, half of them
for the solution and half of them for the uncertainty. The 
training epochs exploiting residual and labeled data use slightly different
losses.

We note that a more general description of the network uncertainties
can be provided by Bayesian neural networks~\cite{Gal2016UncertaintyID,Bollweg:2019skg,Kasieczka:2020vlh,Bellagente:2021yyh,Butter:2021csz}.  We know that their
aleatoric uncertainty, in physics terms essentially the statistical
uncertainty from the training data, can be modeled using Bayesian
neural networks~\cite{Yang_2021}.
However, for our toy example
we find that Bayesian networks require significantly more
training data than the heteroscedastic loss, so we skip them and
instead move on to a different, new method.

%%%%%%%%%%%%%%%%%%%%%%%%%%%%%%%%%%%%%%%%%%%%%%%%
\subsection{Repulsive ensembles}
\label{sec:pinn_repuls}

An alternative way to compute the uncertainty on a network output is
ensembles, provided we ensure that the uncertainty really
covers the probability distribution over the space of 
network functions.
The derivation of repulsive
ensembles~\cite{DAngelo2021RepulsiveDE,Plehn:2022ftl} starts with the usual update
rule minimizing the 
log-probability $p(\theta^t|x_\text{train})$ by gradient descent.

The update rule will be extended to an ensemble of networks,
and its 
coverage of the network space can then be improved by a repulsive
interaction in the update rule.  Such an interaction should take into
account the proximity of the ensemble member $\theta$ to all other
members. We introduce a kernel $k(\theta,\theta_j)$ and add the
interactions with all other weight configurations
\begin{align}
\theta^{t+1} = \theta^t + \alpha 
\nabla_{\theta^t} \left[ \log p(\theta^t|x_\text{train})
- \sum_j  k(\theta^t,\theta^t_j) \right] \; .
\label{eq:update_rule2}
\end{align}
The task is to make sure that this update rule leads to
ensemble members sampling the weight probability, $\theta \sim
p(\theta|x_\text{train})$.

%%%%%%%%%%%%%%%%%%%%%%%%%%%%%%%%%%%%%%%%%%%%%%%%
\subsubsection*{Weight-space density} 

To ensure this sampling property we relate the update rule, or the
discretized $t$-dependence of a weight vector $w(t)$, to a
time-dependent probability density $\rho(\theta,t)$. Just as in the 
setup of 
conditional flow matching networks~\cite{Plehn:2022ftl}, we
can describe the time evolution of a system, equivalently, through an
ODE or a continuity equation,
\begin{align}
\frac{\dd \theta}{\dd t} = v(\theta,t) \qquad \text{or} \qquad 
\frac{\partial \rho(\theta,t)}{\partial t} 
= - \nabla_\theta \left[\upsilon(\theta,t) \rho(\theta,t) \right] \; .
\label{eq:ode}
\end{align}
For a given velocity field $\upsilon(\theta,t)$ the individual paths
$\theta(t)$ describe the evolving density $\rho(\theta,t)$ and the two
conditions are equivalent.  If we choose the velocity field as
\begin{align}
\upsilon(\theta,t) = - \nabla_\theta \log \frac{\rho(\theta,t)}{\pi(\theta)} \; ,
\end{align}
these two equivalent conditions read
\begin{align}
\frac{\dd \theta}{\dd t} &= - \nabla_\theta \log \frac{\rho(\theta,t)}{\pi(\theta)} 
\notag \\
\frac{\partial \rho(\theta,t)}{\partial t} 
%&= \nabla_\theta \left[ \rho(\theta,t) \nabla_\theta \log \frac{\rho(\theta,t)}{\pi(\theta)}\right] \notag \\
&= - \nabla_\theta \left[ \rho(\theta,t) \nabla_\theta \log \pi(\theta) \right]
+ \nabla_\theta^2 \log \rho(\theta,t) \; .
\label{eq:fp}
\end{align}
The continuity equation becomes the Fokker-Planck equation, for which
$ \rho(\theta,t) \to \pi(\theta)$ is the unique stationary probability
distribution.

Next, we relate the ODE in Eq.\eqref{eq:fp} to the update rule for
repulsive ensembles, Eq.\eqref{eq:update_rule2}.  The discretized
version of the ODE is
\begin{align}
\frac{\theta^{t+1}-\theta^t}{\alpha}
= - \nabla_{\theta^t} \log \frac{\rho(\theta^t)}{\pi(\theta^t)} \; .
\end{align}
If we do not know the density $\rho(\theta^t)$ explicitly, we can
approximate it as a superposition of kernels,
\begin{align}
\rho(\theta^t) \approx \frac{1}{n} \sum_{i=1}^n k(\theta^t,\theta_i^t)
\qquad \text{with} \qquad 
\int d\theta^t \rho(\theta^t) 
%&= \frac{1}{n} \sum_{i=1}^n \int d\theta^t k(\theta^t,\theta_i^t) \notag \\
%&= \frac{1}{n} \sum_{i=1}^n 1 
= 1 \; .
\label{eq:normalize}
\end{align}
We can insert this kernel approximation into the discretized ODE and find
\begin{align}
\frac{\theta^{t+1}-\theta^t}{\alpha}
%&= - \nabla_{\theta^t} \left[ \log \rho(\theta^t) - \log \pi(\theta^t) \right]
%\notag \\
%&= \nabla_{\theta^t} \log \pi(\theta^t)
%- \nabla_{\theta^t} \log \left[ \frac{1}{n} \sum_i k(\theta^t,\theta_i^t) %\right]
%\notag \\
%&= \nabla_{\theta^t} \log \pi(\theta^t)
%- \nabla_{\theta^t} \log \sum_i k(\theta^t,\theta_i^t) 
%\notag \\
&= \nabla_{\theta^t} \log \pi(\theta^t)
- \frac{\nabla_{\theta^t} \sum_i k(\theta^t,\theta_i^t)}{\sum_i k(\theta^t,\theta_i^t)}
\label{eq:fp_update}
\end{align}
This form can be identified with the update rule in 
Eq.\eqref{eq:update_rule2} by setting 
%.  To make them identical, we first identify
%
%\begin{align}
$\pi(\theta) \equiv p(\theta|x_\text{train})$, 
which means that the update rule will converge to the 
correct probability.
Second, we add the normalization
term of Eq.\eqref{eq:fp_update} to our original kernel in
Eq.\eqref{eq:update_rule2},
\begin{align}
\nabla_{\theta^t} \sum_i k(\theta^t,\theta_i^t)
\; \to \; 
\frac{\nabla_{\theta^t} \sum_i k(\theta^t,\theta_i^t)}{\sum_i k(\theta^t,\theta_i^t)} \; ,
\label{eq:normalized_grad}
\end{align}
to ensure that the update rule with an appropriate kernel 
leads to the correct density.

%%%%%%%%%%%%%%%%%%%%%%%%%%%%%%%%%%%%%%%%%%%%%%%%
\subsubsection*{Function-space density}

So far, we consider ensembles with a repulsive force in weight
space. However, we are interested in the function the network encodes
and not the latent or weight representation.  For instance, two
networks encoding the same function could be constructed by permuting
the weights of the hidden layers, unaffected by a repulsive force in
weight space.  This is why we prefer a repulsive force in the space of
network outputs $f_\theta(x)$.

Symbolically, we can then write the update rule from
Eq.\eqref{eq:update_rule2} with the normalization of
Eq.\eqref{eq:normalized_grad} as
\begin{align}
\frac{f^{t+1} - f^t}{\alpha} = 
\nabla_{f^t} \log p(f|x_\text{train})
- \frac{\sum_j \nabla_{f^t} k(f,f_j)}{\sum_j k(f,f_j)} \; .
\label{eq:update_rule_fspace}
\end{align}
The network training is still defined in weight space, so we have to
translate the function-space update rule into weight space using the
appropriate Jacobian
\begin{align}
\frac{\theta^{t+1}-\theta^t}{\alpha} 
%&= \frac{\partial f^t}{\partial \theta^t} 
%\left[ \nabla_{f^t} \log p(f_{\theta^t}|x_\text{train})
%- \frac{\sum_j \nabla_{f} k(f_{\theta^t},f_{\theta_j^t})}{\sum_j %k(f_{\theta^t},f_{\theta_j^t})} \right] \notag \\
&= 
\nabla_{\theta^t} \log p(\theta^t|x_\text{train})
- \frac{\partial f^t}{\partial \theta^t} \frac{\sum_j \nabla_{f} k(f_{\theta^t},f_{\theta_j^t})}{\sum_j k(f_{\theta^t},f_{\theta_j^t})} \; .
\end{align}
Furthermore, we cannot evaluate the repulsive kernel in function
space, so we have to evaluate the function for a finite batch of
points $x$,
\begin{align}
\frac{\theta^{t+1} - \theta^t}{\alpha} 
\approx \nabla_{\theta^t} \log p(\theta^t|x_\text{train})
- \frac{\sum_j \nabla_{\theta^t} k(f_{\theta^t}(x),f_{\theta_j^t}(x))}{\sum_j k(f_{\theta^t}(x),f_{\theta_j^t}(x))}  \; .
\label{eq:update_rule_fspace2}
\end{align}
%

%%%%%%%%%%%%%%%%%%%%%%%%%%%%%%%%%%%%%%%%%%%%%%%%
\subsubsection*{Loss function} 

Finally, we turn the update rule in Eq.\eqref{eq:update_rule_fspace2}
into a loss function for the repulsive ensemble training.  We transform the
probability into
a tractable likelihood loss with a Gaussian prior,
\begin{align}
    \log p(\theta|x_\text{train}) 
    = \log p(x_\text{train}|\theta) \; 
    - \frac{|\theta|^2}{2 \sigma^2} + \text{const} \; .
\end{align}
Given a training dataset of size $N$, we evaluate the likelihood on
batches of size $B$, so Eq.\eqref{eq:update_rule_fspace} becomes
\begin{align}
    \frac{\theta^{t+1} - \theta^t}{\alpha}
\approx \nabla_{\theta^t} \frac{N}{B} \sum_{b=1}^B \log p(x_b|\theta)
- \frac{\sum_j \nabla_{\theta^t} k(f_{\theta^t}(x),f_{\theta_j^t}(x))}{\sum_j k(f_{\theta^t}(x),f_{\theta_j^t}(x))} 
- \nabla_{\theta^t} \frac{|\theta|^2}{2 \sigma^2}
\; .
\end{align}
Here, $f_{\theta^t}(x)$ is to be understood as evaluating the function
for all samples $x_1,\dots,x_B$ in the batch.

To turn the update rule into a loss function, we flip the sign of term
in the gradient, divide it by $N$ to remove the scaling with the size
of the training dataset, and sum over all members of the
ensemble. Since the gradients of the loss function are computed with
respect to the parameters of all networks in the ensemble, we need to
ensure the correct gradients of the repulsive term using a
stop-gradient operation, denoted with an overline
$\overline{f_{\theta_j}(x)}$. The loss function for repulsive
ensembles then reads
\begin{align}
    \loss = \sum_{i=1}^n \left[- \frac{1}{B} \sum_{b=1}^B \log p(x_b|\theta_i)
    + \frac{1}{N} \frac{\sum_{j=1}^n k(f_{\theta_i}(x), \overline{f_{\theta_j}(x)})}{\sum_{j=1}^n k(\overline{f_{\theta_i}(x)}, \overline{f_{\theta_j}(x)})} + \frac{|\theta_i|^2}{2N \sigma^2}  \right] \; .
\label{eq:loss_re}
\end{align}
The prior has just become an L2-regularization with prefactor $1/(2N\sigma^2)$, 
like in a Bayesian neural network.

%%%%%%%%%%%%%%%%%%%%%%%%%%%%%%%%%%%%%%%%%%%%%%%%
\subsubsection*{Kernel in function space}

A typical choice for the Kernel introduced in Eq.\eqref{eq:normalize}
is a Gaussian. For the loss in Eq.\eqref{eq:loss_re} this has to be a
Gaussian in the multi-dimensional function space, evaluated over a
sample,
\begin{align}
k(f_{\theta_i}(x),f_{\theta_j}(x)) =
\prod_{b=1}^B \exp \left( - \frac{|f_{\theta_i}(x_b) - f_{\theta_j}(x_b)|^2}{h} \right) \; .
\label{eq:gaussian_kernel}
\end{align}
The width $h$ should be chosen such that the width of the distribution
is not overestimated while still ensuring that it is sufficiently
smooth. This can be achieved with the median
heuristic~\cite{liu2016stein},
\begin{align}
    h &= \frac{\text{median}_{ij}\left(
    \sum_b |f_{\theta_i}(x_b) - f_{\theta_j}(x_b)|^2
    \right)}{2 \log(n + 1)} \; ,
\end{align}
with the number of ensemble members $n$.

%%%%%%%%%%%%%%%%%%%%%%%%%%%%%%%%%%%%%%%%%%%%%%%%
\subsection{Uncertainties}
\label{sec:pinn_res}

To show that we can describe uncertainties of PINNs using a
heteroscedastic loss and repulsive ensembles, we use the harmonic
oscillator toy model from Sec.~\ref{sec:pinn_harmonic}. The only
difference is that we, for instance, distribute the labeled data points
such that they become sparse for late times, to see if we can track
this statistic uncertainty in the training data in the uncertainty of
the network output.

%%%%%%%%%%%%%%%%%%%%%%%%%%%%%%%%%%%%%%%%%%%%%%%%
\subsubsection*{Sparse and stochastic data}

\begin{figure}[t]
  \includegraphics[width = 0.495\textwidth]{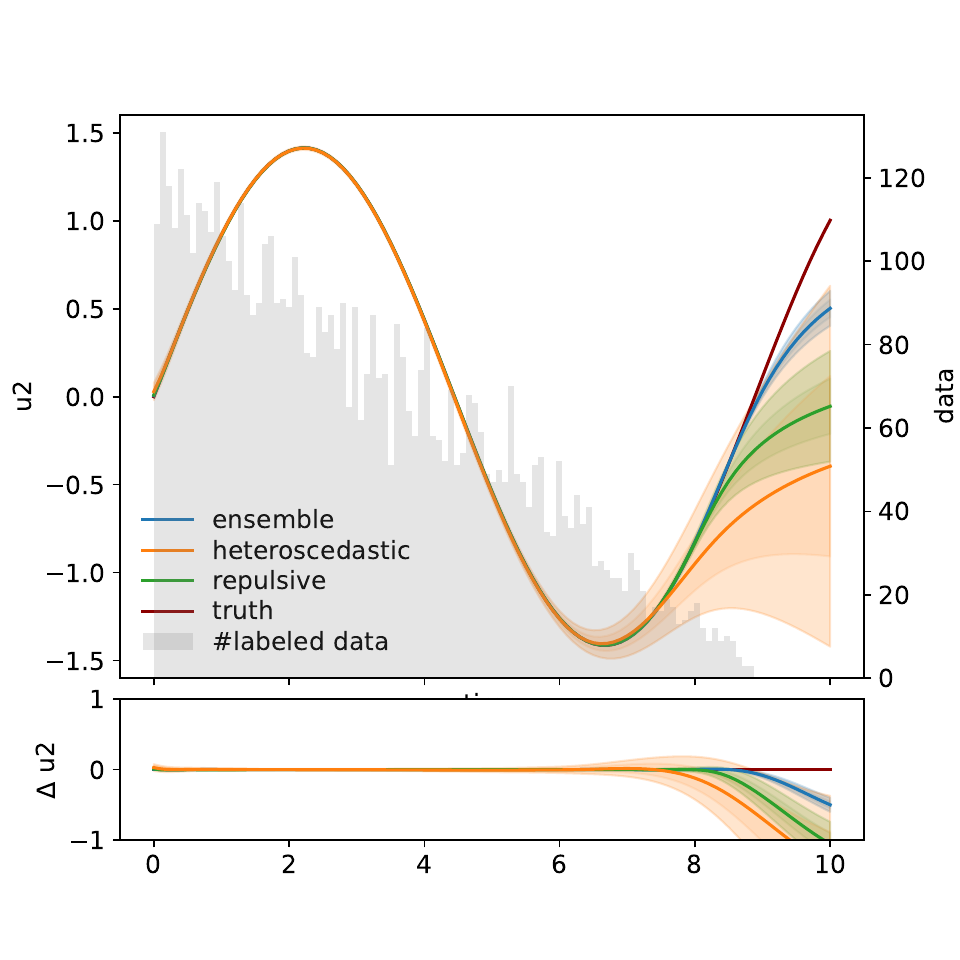}
  \includegraphics[width = 0.495\textwidth]{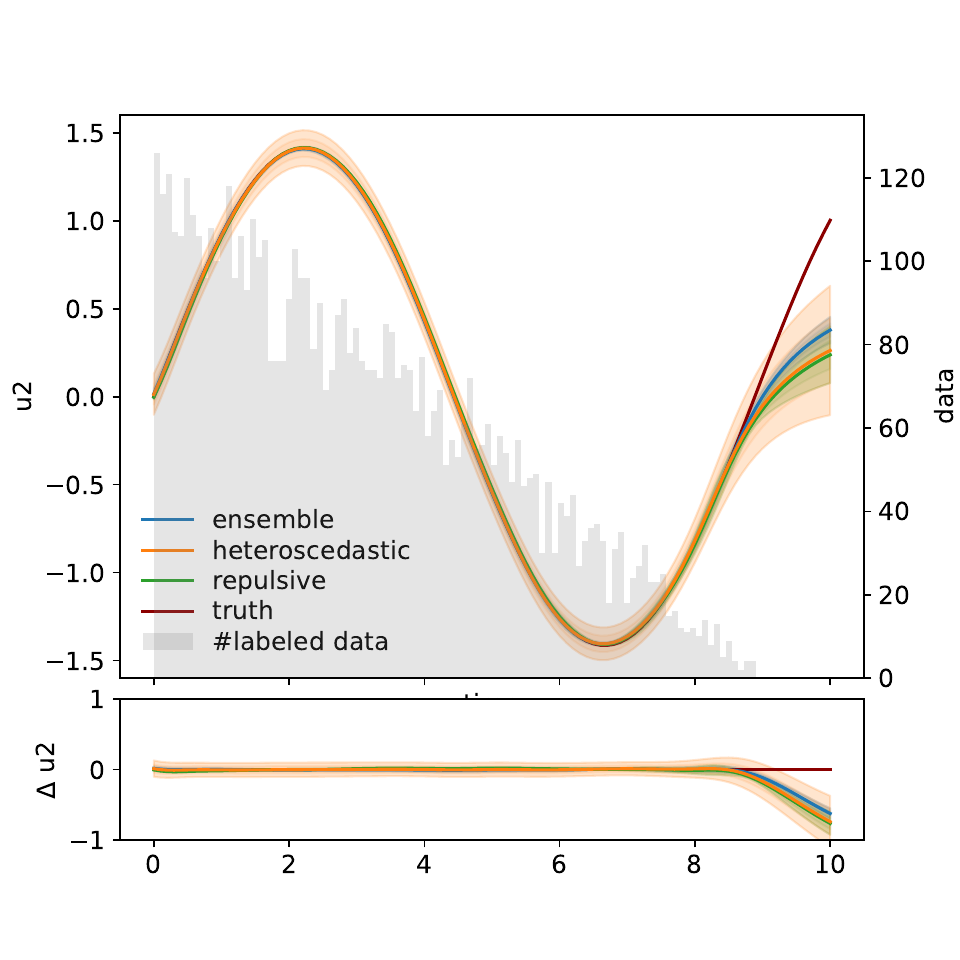}
  \caption{Learned harmonic oscillator with sparse training data at
    late times. For the training we only use labeled data points, 
    defining a simple regression task. In
    the left panel the training data is exact, in the right panel it
    includes noise. The error bars correspond to 68\% and 95\%~CL.}
 \label{fig:ho_stoch}
\end{figure}

First, we look at the trained network and its uncertainty estimate if
we only include labeled data points and reduce the density of training
data towards late times. We can do this without and with noise in the
labeled data. This way, the training has no access to the late-time
regime. In our setup the labels $u$ and $\dot{u}$ are separate, so
this network training also misses all information about the
differential equation.  The decreasing distribution of labeled data
points is given in the background histogram of
Fig.~\ref{fig:ho_stoch}, creating a smooth extrapolation problem
towards late times for a simple regression.

The left panel demonstrates the effect of increasingly sparse data
without noise.  Indeed, the uncertainty increases with time, as the
density of labeled training points
decreases~\cite{seitzer2022on}. Both, the repulsive ensemble and the
heteroscedastic network deviate from the true solution for $t>8$,
which means they have learned the shape of the minimum even though
there is very little data beyond $t=6$.  The repulsive ensemble
remains more stable than the heteroscedastic network, which can be
explained by the stabilizing effect of ensembling. For both, the
heteroscedastic network and the repulsive ensembles, the error bar
increases fast enough to cover the deviation from the true solution to
$t=9$. Beyond this point the error bar is not conservative in covering
the uncertainty related to missing training data altogether. The
classic ensemble without repulsive term happens to guess the correct
solution reasonably well, but without a meaningful spread.

%-----------------------------------------------
%-----------------------------------------------

In the right panel of Fig.~\ref{fig:ho_stoch} we see what happens when
we switch to noisy data. Now, the labeled data points still encode the
details of the differential equation, but with Gaussian noise on the
$u$ and $\dot u$ information of mean zero and width $0.1$. The
heteroscedastic network captures this stochasticity as an additional
source of uncertainty over the entire time range, while the members of
the (repulsive) ensemble each determine the best solution without a
visible spread. On the other hand, in this case it is not clear how
useful the heteroscedastic uncertainty is, given that the noisy data
does allow all networks to learn the true distributions very well. At
late times, the noise has a counter-intuitive effect on the
extrapolation; all predictions become better, and the the reduced
uncertainties confirm this trend. The central values and the error
bars for the heteroscedastic network and the repulsive ensembles loose
all their reliability in the region without data, $t>9$.

The simple bottom line of both tests, with and without noise, is that
extrapolation for a simple regression task works as long as there is
some data, and beyond this point it fails. This is true for the
central value learned by the network and for the uncertainty
estimate. We emphasize that in many practical applications for
instance in particle physics this uncertainty estimate can still be
used, because out-of-distribution data is just a limit of increasingly
sparse training data.

%%%%%%%%%%%%%%%%%%%%%%%%%%%%%%%%%%%%%%%%%%%%%%%%
\subsubsection*{ODE extrapolation}

%-----------------------------------------------
\begin{figure}[t]
  \includegraphics[width = 0.495\textwidth]{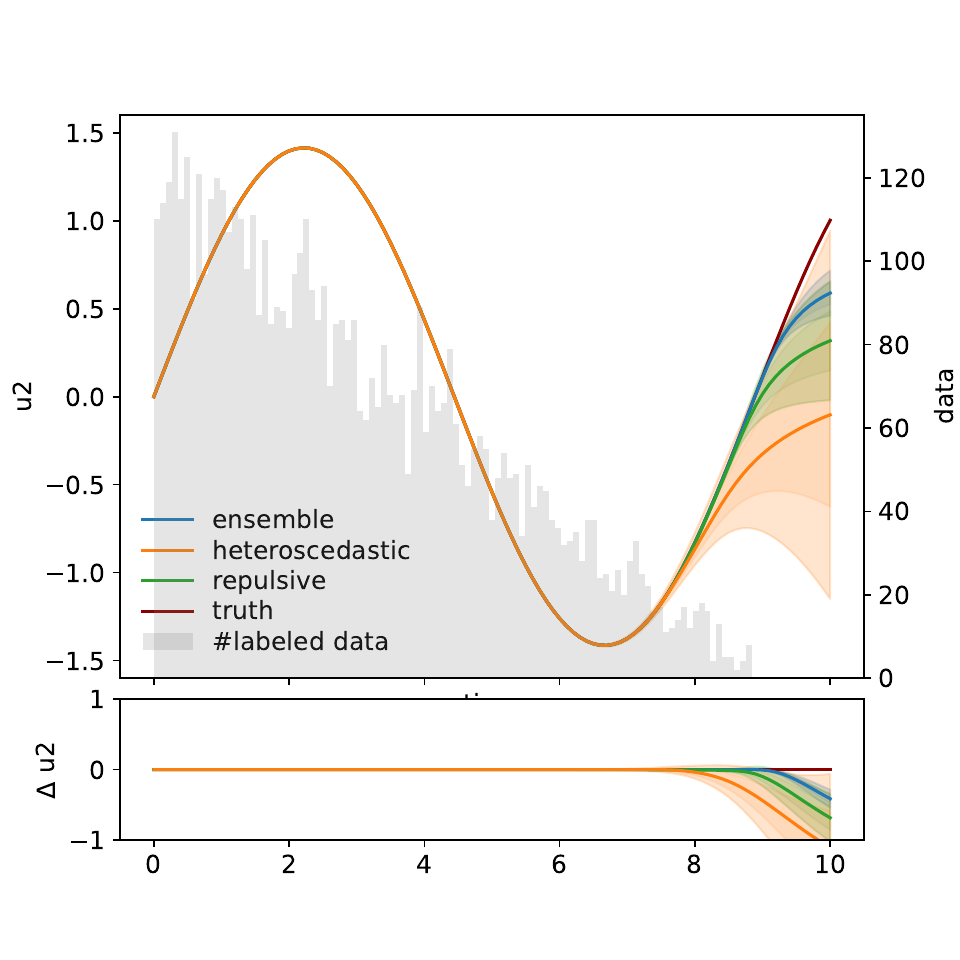}
  \includegraphics[width = 0.495\textwidth]{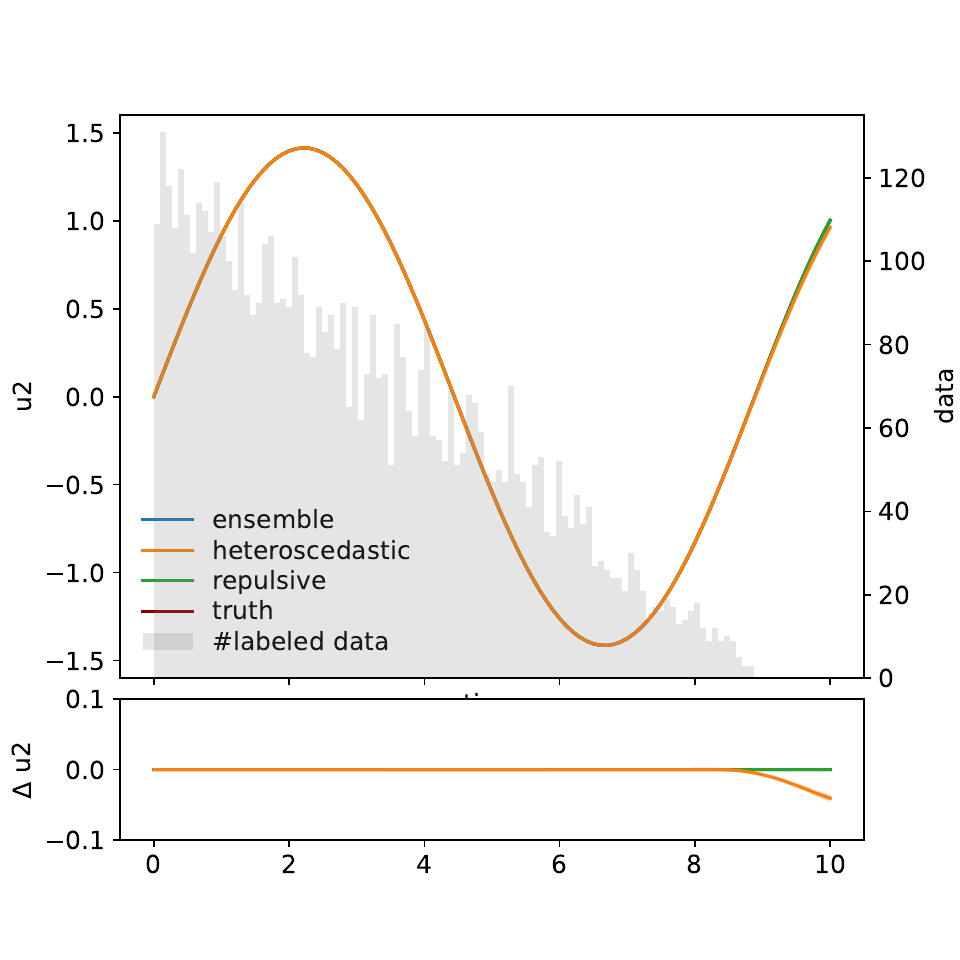}
 \caption{Learned harmonic oscillator adding the ODE loss enforcing
   the differential equation.  For the left panel the additional 
   residual points are distributed like the labeled point, for the right 
   panels we add 10000 residual point uniformly over time. The
   error bars correspond to 68\% and 95\%~CL.}
 \label{fig:ho_ode}
\end{figure} 
%-----------------------------------------------

The strength of PINNs is that they can extrapolate to regions without
labeled data, using additional residual data trained with the ODE loss.
At these points the network can confirm that its output fulfills the
differential equation. We train with the two datasets alternatingly,
one epoch using the labeled data point and one epoch using residual
points, both computing the loss in Eq.\eqref{eq:heteroscedasticPINN}.

First, we include residual data with the same time distribution as the
labeled data. In practice, we strip the labeled data of the additional
information and add the remaining $t$-values as residual point.  In
the left panel of Fig.~\ref{fig:ho_ode} we see that the PINNs become
slightly more accurate at large times than they are in
Fig.~\ref{fig:ho_stoch}.  This is true at least in the case without
noise, while we have checked that the improvement is not visible for
noisy data. This means that with labeled and residual training data
covering the same time range, the network does not learn the
differential distributions precisely enough to provide a reliable
description towards late times. The learned network uncertainty
confirms the behavior of the central prediction.

Second, we add 10000 residual training points equally distributed over
time.  Without noise, these models reproduce the true function
extremely well, over the entire time range and with correspondingly
small uncertainties from the heteroscedastic loss as well as the
repulsive ensembles.  The uniformly distributed residual points leave
no wiggle room to the network training anymore and learn the full time
range. We note, however, that this does not count as an extrapolation,
because the residual data also covers the entire range.

%%%%%%%%%%%%%%%%%%%%%%%%%%%%%%%%%%%%%%%%%%%%%%%%
\subsubsection*{Interpolation turning extrapolation}

%-----------------------------------------------
\begin{figure}[t]
  \includegraphics[width = 0.495\textwidth]{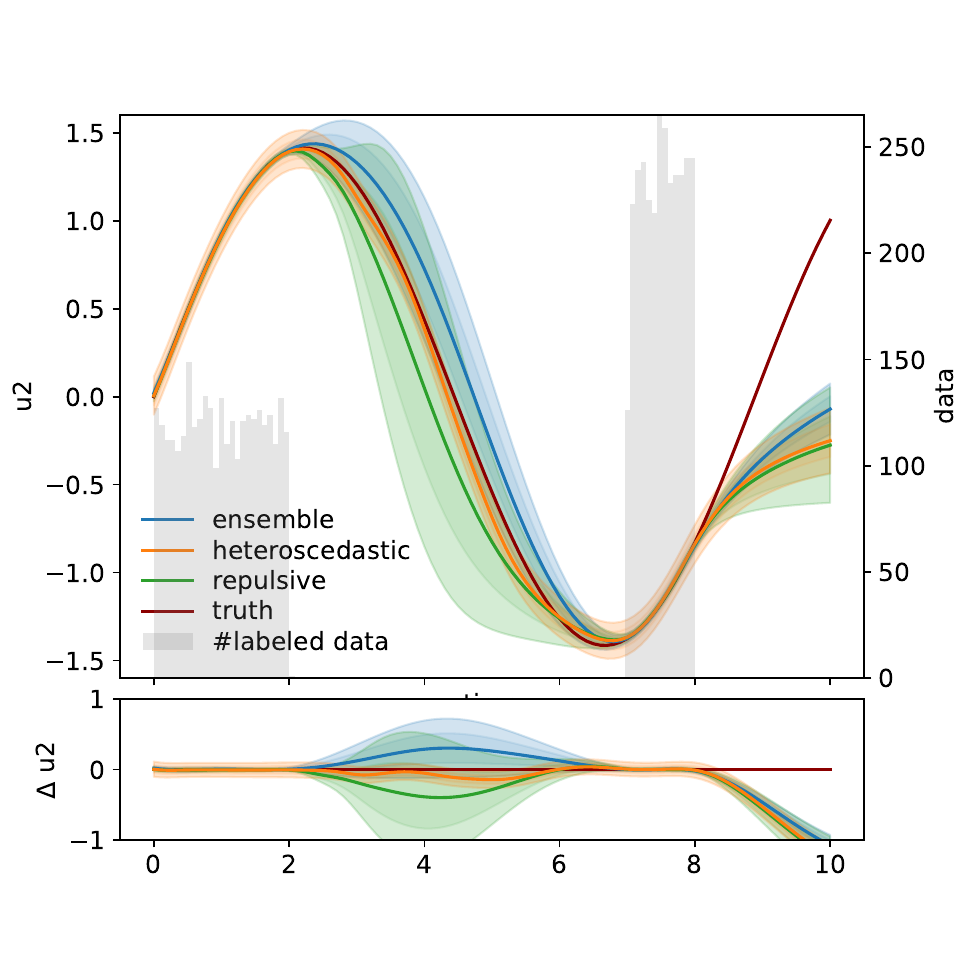}
  \includegraphics[width = 0.495\textwidth]{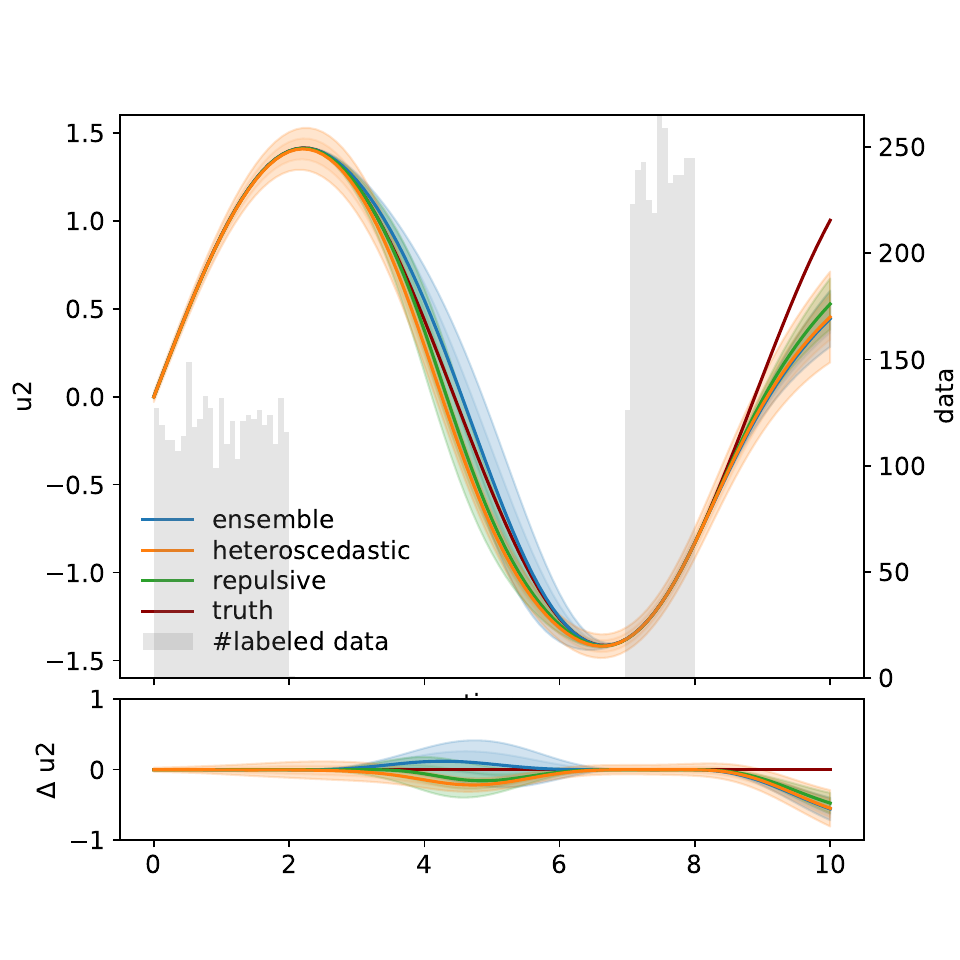}
  \caption{Learned harmonic oscillator with split training data and no
    noise. In the left panel we only use the labeled training point,
    in the right panel we add residual points distributed the same way
    as the labeled points. The error bars correspond
    to 68\% and 95\%~CL.}
  \label{fig:ho_inter}
\end{figure} 
%-----------------------------------------------

In a sufficiently high number of dimensions, even an apparent
interpolation relies on such a low density of training data that it
resembles the typical extrapolation illustrated in
Fig.~\ref{fig:ho_stoch}. For the extrapolation we know that both, the
heteroscedastic loss and the repulsive ensembles assign an increasing
error bar towards the data-deprived region, with a conservative
uncertainty estimate for as long as there is training data. The
question is if the same happens for a wide interpolation.

For an extrapolation-like interpolation we assume training data
at small times, $t=0 < 2$, and $t=7~...~8$. This forces the network to
interpolate over a large time window and extrapolate to very late times.
In the left panel of Fig.~\ref{fig:ho_inter} see that the wide
interpolation challenges the usual ensemble of networks, indicated by
the poor agreement with the true solution. The spread of the classic
ensemble barely covers the difference from the truth.  The situation
improves with repulsive ensembles, which provide a better central
prediction and much more conservative error bars in both sparse
regions. Especially for the late times we see that the uncertainty
assigned but the repulsive ensembles covers the deviation from the
true solution well.  In the interpolation region the heteroscedastic
network covers a much smaller family of functions. While the central
value deviates from the true solution at a similar level as the
repulsive ensembles, the error bar is smaller and not really
conservative for the extrapolation.

In the right panel of Fig.~\ref{fig:ho_inter}, we again add residual
data following the same distribution as the labeled data. This means
the network can learn the differential equation using the ODE
loss. From Fig.~\ref{fig:ho_ode} (left) we know that this has hardly any 
effect on regions with enough data or on actual extrapolation.
However, here we see that the residual data and the ODE loss have
a significant effect on the uncertainty estimate for the wide interpolation.

As a final remark --- given that we know that neural
networks are extremely good at interpolating, the question becomes
what we expect from an error bar in the interpolation region. Either
we argue that the network should consider a wide interpolation
an extrapolation and admit that there is not enough data to capture 
possible features in the sparsely probed region. In that case the error bar 
should be large. Or we trust the network to interpolate well, under the 
assumption that there are no additional features, in which case a small
uncertainty reflects the confidence of the network training.

%%%%%%%%%%%%%%%%%%%%%%%%%%%%%%%%%%%%%%%%%%%%%%%%
\section{Supernova PINNulator}
\label{sect:PINNasEmulator}

The computation of the distance moduli $\mu$ of the type Ia supernovae make
a compelling use case for PINNs in cosmology. For a known Hubble
function the luminosity distances are computed through integration, or
equivalently by solving a simple differential equation
\begin{align}
  \mu = 5\log_{10} d_L(z,\lambda) + 10 
  \quad\text{with}\qquad
  d_L(z,\lambda)
  = (1+z) \, c \int_0^z \dd{z^\prime} \frac{1}{{H(z^\prime,\lambda)}} \; .
	%d_L(z) = (1+z) \, \chi_H \int_{0}^{z}\dd{z^\prime}\:
	%\frac{1}{\sqrt{\Omega_m (1+z^\prime)^3 + (1-\Omega_m) (1+z^\prime)^{3(1+\omega)}}}.
\label{eq:DistanceModulus}
\end{align}
Depending on the assumptions on the relevant components of the universe, we 
can follow different strategies. In this section we focus on a two-fluid 
cosmology including dark matter and dark energy, $w$CDM, assuming a constant 
$w(z) < -1/3$ to ensure accelerated expansion. The cosmologicial constant 
$\Lambda$ with $w=-1$ is a particular sub-case. If we only assume the 
FLRW-symmetries, the Hubble function $H(z)$ can take any form allowed by the 
data. We come back to this second option in Sec.~\ref{sect:PINNference}.

%%%%%%%%%%%%%%%%%%%%%%%%%%%%%%%%%%%%%%%%%%%%%%%%
\subsubsection*{Luminosity-distance PINN}

PINNs can learn to predict luminosity distances~\cite{10.1093/mnras/Rover} and 
hence distance moduli, when trained to emulate the solution of the differential 
equation as a function of redshift. The resulting emulator, approximating the
distance moduli as in Eq.\eqref{eq:DistanceModulus}, can be used to
speed up inference. 

The Hubble function depends on a set of cosmological parameters $\lambda$, and we carry this additional argument through our derivation.  To apply PINNs as illustrated in Eq.\eqref{eq:generalODE}, the luminosity distance is expressed through the ODE
\begin{align}
	\frac{\dd \tilde{d}_L(z,\lambda)}{\dd z}- \frac{\tilde{d}_L(z,\lambda)}{1+z} - \frac{1+z}{\tilde{H}(z,\lambda)} =0
    \qquad \text{with} \qquad d_L(0,\lambda) = 0 \; . 
    \label{eq:lumiODE}
\end{align}
Here, $\tilde{d}_L = d_L H_0/c$ and $\tilde{H}(z,\lambda) =
H(z,\lambda)/H_0$ are de-dimensionalized and ensure solutions of order
unity. This makes PINN training more stable~\cite{Wang2023AnEG}.
To learn the solution to Eq.\eqref{eq:lumiODE}, we choose the
cosmological parameters and the functional form for the Hubble
function to conform to a flat two-fluid cosmology where the dark energy component 
has a constant equation of state $w$, similar Ref.~\cite{Chantada:2022bdf},
\begin{align}
    \frac{H(z,\lambda)}{H_0} 
    = \sqrt{\Omega_m (1+z)^3 + (1-\Omega_m) (1+z)^{3(1+w)}} \; .
\label{eq:hubbleParametrized}
\end{align}
Our PINN tracks three cosmological input parameters denoted as $\lambda$: 
(i) the redshift $z$; (ii) the energy density of matter $\Omega_m$;
and {iii) the dark energy equation
of state parameter $w$. In this subsection, we fix the Hubble
parameter to $70~\text{km}/\text{s}/\text{Mpc}$.

The two relevant losses defined in~Eq.\eqref{eq:generalLoss} can be read off Eq.\eqref{eq:lumiODE}
\begin{align}
	\loss_\text{IC} 
 &= \frac{1}{N}\sum_{i=0}^{N}\left[ d_{L,\theta}(0,\lambda_i) \right] ^2 \notag \\
 \loss_\text{ODE} 
 &= \frac{1}{N}\sum_{i=0}^{N}\left[ \frac{\dd d_{L,\theta}(z_i,\lambda_i)}{\dd z}- \frac{d_{L,\theta}(z_i,\lambda_i)}{1+z_i} - \frac{1+z_i}{H(z_i,\lambda_i)}\right]^2 \; .
\end{align}
The index $i$ counts $N$ elements $(z,
\lambda)_i$, generated uniformly over the relevant parameter
ranges.

As in the toy example, we construct heteroscedastic versions of the
MSE losses to learn the uncertainties from the training data,
\begin{align}
    \label{eq:supernovaHeteroscedasticExplicit}
    \loss_\text{IC,het} 
 &= \frac{1}{N}\sum_{i=0}^N \left[
 \frac{d_{L,\theta}(0,\lambda_i)^2}{2 \sigma_\theta(0,\lambda_i)^2}
 + \log\sigma_\theta(0,\lambda_i)
 \right] \\
    \loss_\text{ODE,het} 
    &= \frac{1}{N}\sum_{i=1}^N \left[
    \frac{1}{2 \sigma_\theta(z_i,\lambda_i)^2}
    \left(
    \frac{\dd d_{L,\theta}(z_i,\lambda_i)}{\dd z}
    - \frac{d_{L,\theta}(z_i,\lambda_i)}{1+z_i} 
    - \frac{1+z_i}{H(z_i,\lambda_i)}\right)^2
    + \log\sigma_\theta(z_i,\lambda_i)
    \right] \; . \notag 
\end{align} 
Our small network uses five hidden layers with 100 nodes each. The
one-dimensional output approximates the luminosity distance.  The
$10^5$ residual training points are generated uniformly in the ranges
$z \in [0,1.8]$, $\Omega_m \in [0,1]$, and $w \in [-1.6,-0.5]$.  We
will see that the network training is good enough that we do not have
to consider labeled data for the PNN emulator.  A similar model was
used as an emulator in Ref.~\cite{10.1093/mnras/Rover}, to constrain
the matter density and the equation of state parameter using the
Union2.1-dataset~\cite{suzuki_hubble_2012, amanullah_spectra_2010,
  kowalski_improved_2008}.

%-----------------------------------------------
\begin{figure}[b!]
  \includegraphics[width = 0.495\textwidth]{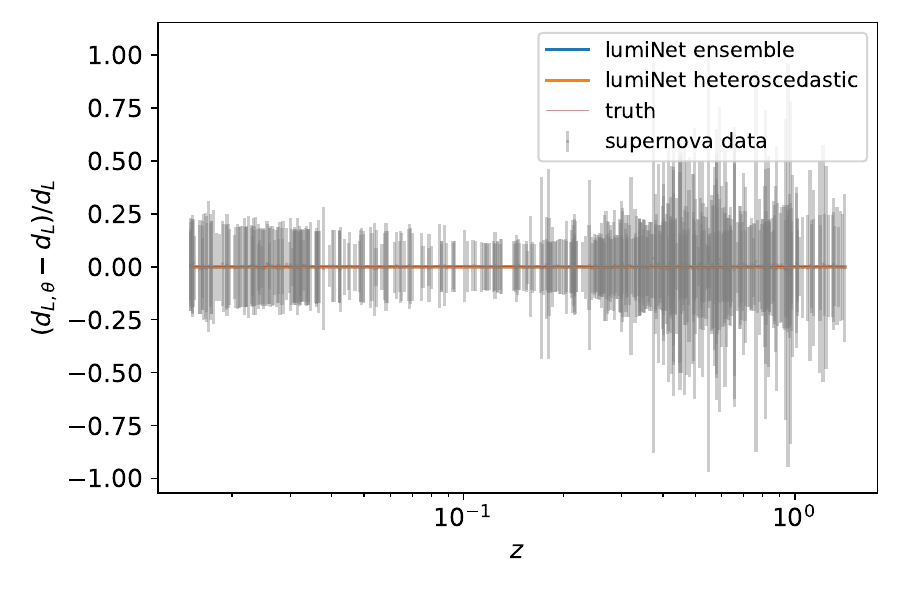}
  \includegraphics[width = 0.495\textwidth]{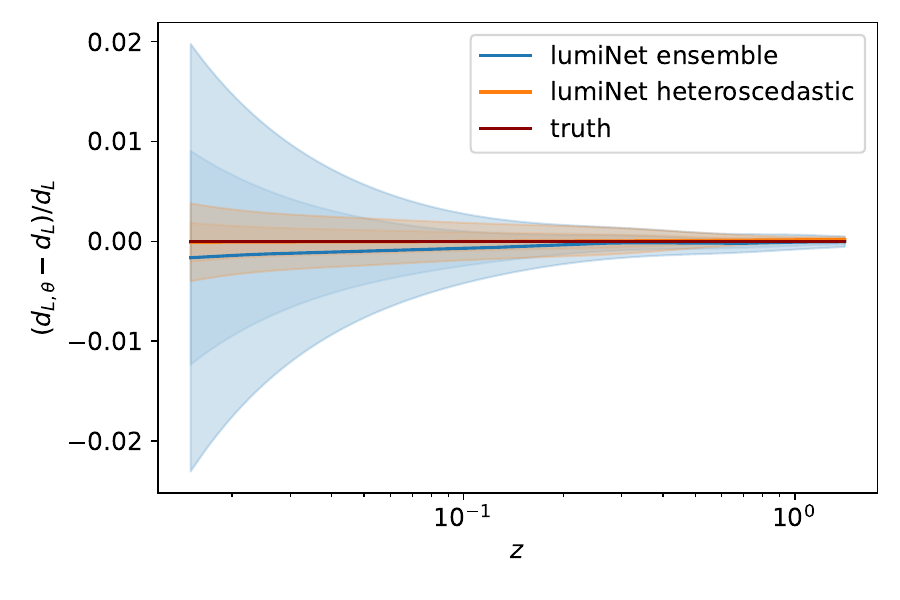}
	\caption{Learned luminosity distance from residual points only.
 The left panel compares the heteroscedastic PINN uncertainty to 
 the experimental uncertainties in the Union2.1 dataset. The right panel shows the relative difference between the learned and true solutions. For the ensemble spread we train 10 independent models on different
 data points. }
 \label{fig:supernovaEmulatorErrorbars}
\end{figure}
%-----------------------------------------------

%%%%%%%%%%%%%%%%%%%%%%%%%%%%%%%%%%%%%%%%%%%%%%%%
\subsubsection*{Luminosity-distance emulator}

Figure~\ref{fig:supernovaEmulatorErrorbars} demonstrates the accuracy
of the PINN emulator assuming the best fit parameters of the Union2.1
dataset. For this parameter choice, the left panel shows that the
heteroscedastic uncertainty on the trained PINN emulators are more
than an order of magnitude smaller than the experimental
uncertainties. The right panel shows that the spread of ten MSE
trained PINNs is larger than the uncertainty estimation obtained when
training with a heteroscedastic loss.

We can understand this behavior from the training.  If we only rely on
residual points, the solution is probed exactly for a given
redshift. The heteroscedastic error will not be affected by
stochasticity or noise, but capture the limitations from the
expressivity of the neural network.
In addition, the heteroscedastic training doubles the network output
and allows the network to adjust the central prediction and the error
as a function of time.  Rather then trying to adjust a network with
limited expressivity to data of arbitrary precision, it can offload
some problems to the learned uncertainty, which can, and does,
stabilize the training and the ultimate agreement with the true
solution.

To allow for cosmological inference, an emulator of distance moduli
also has to approximate the solution to the differential equation away
from the best fit parameters. To test the reliability of the PINN we
generate 1000 test data points from the same distribution as the
training data. For this test data we first compute the true luminosity
distance using Eq.\eqref{eq:lumiODE}. Then, we generate the learned
luminosity distances and their uncertainties from the PINN. The left
panel of Fig.~\ref{fig:supernovaTestSigmas} shows the deviation of the
PINN prediction from the true solution. The spread of the ensemble
trained with an MSE loss deviates from the truth by less than two
percent.  The heteroscedastic training improves this agreement to
better than one percent.  However, in the right panel of Fig.~\ref{fig:supernovaTestSigmas}, we also see that the relative
uncertainties grow rapidly for small redshifts, because the initial
condition for the luminosity distance is also small. This requires
higher absolute precision at small redshifts.  

Nevertheless, we find
that especially the PINNs trained with the heteroscedastic loss are
extremely precise even without resorting to labeled data
training. This is definitely sufficient to be used as an emulator for
the luminosity distance for the Union2.1 or
Pantheon+~\cite{Scolnic_2022} data, which come with experimental
errors of around $10$\%.

%-----------------------------------------------
\begin{figure}[t]
	\includegraphics[width = 0.495\textwidth]{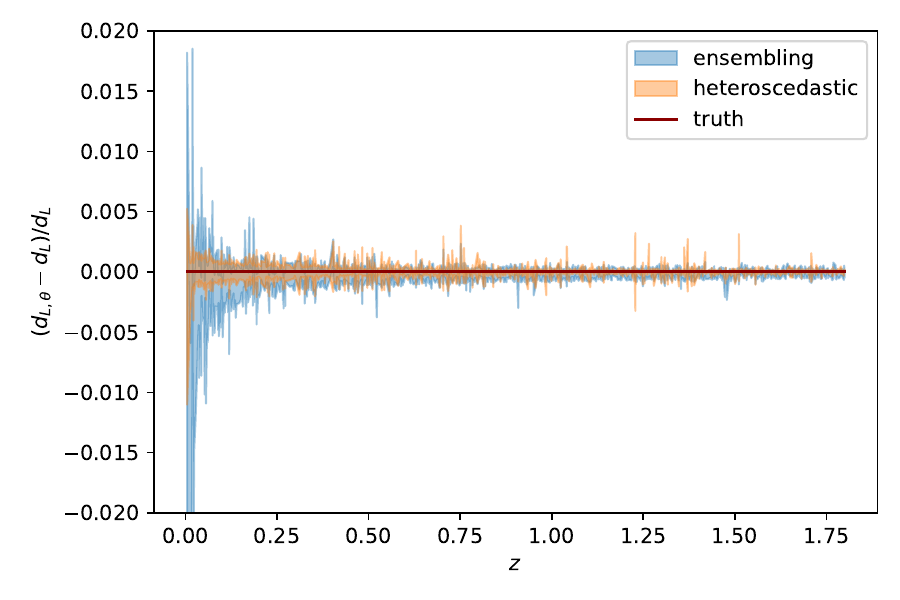}
	\includegraphics[width = 0.495\textwidth]{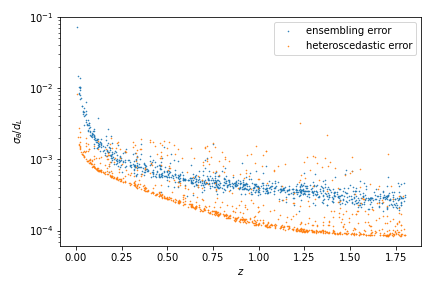}
	\caption{PINN accuracy for data points uniformly sampled from the same cosmological parameter ranges as the training points. The left panel shows  the error bands around the true solution, the right panel the evolution of the ensemble spread and the heteroscedastic uncertainty with redshift.}
 \label{fig:supernovaTestSigmas}
\end{figure} 
%-----------------------------------------------

%%%%%%%%%%%%%%%%%%%%%%%%%%%%%%%%%%%%%%%%%%%%%%%%
\section{Supernova PINNference}
\label{sect:PINNference}

%-----------------------------------------------
\begin{figure}[t]
    \centering
    \includegraphics[width = 0.99\textwidth]{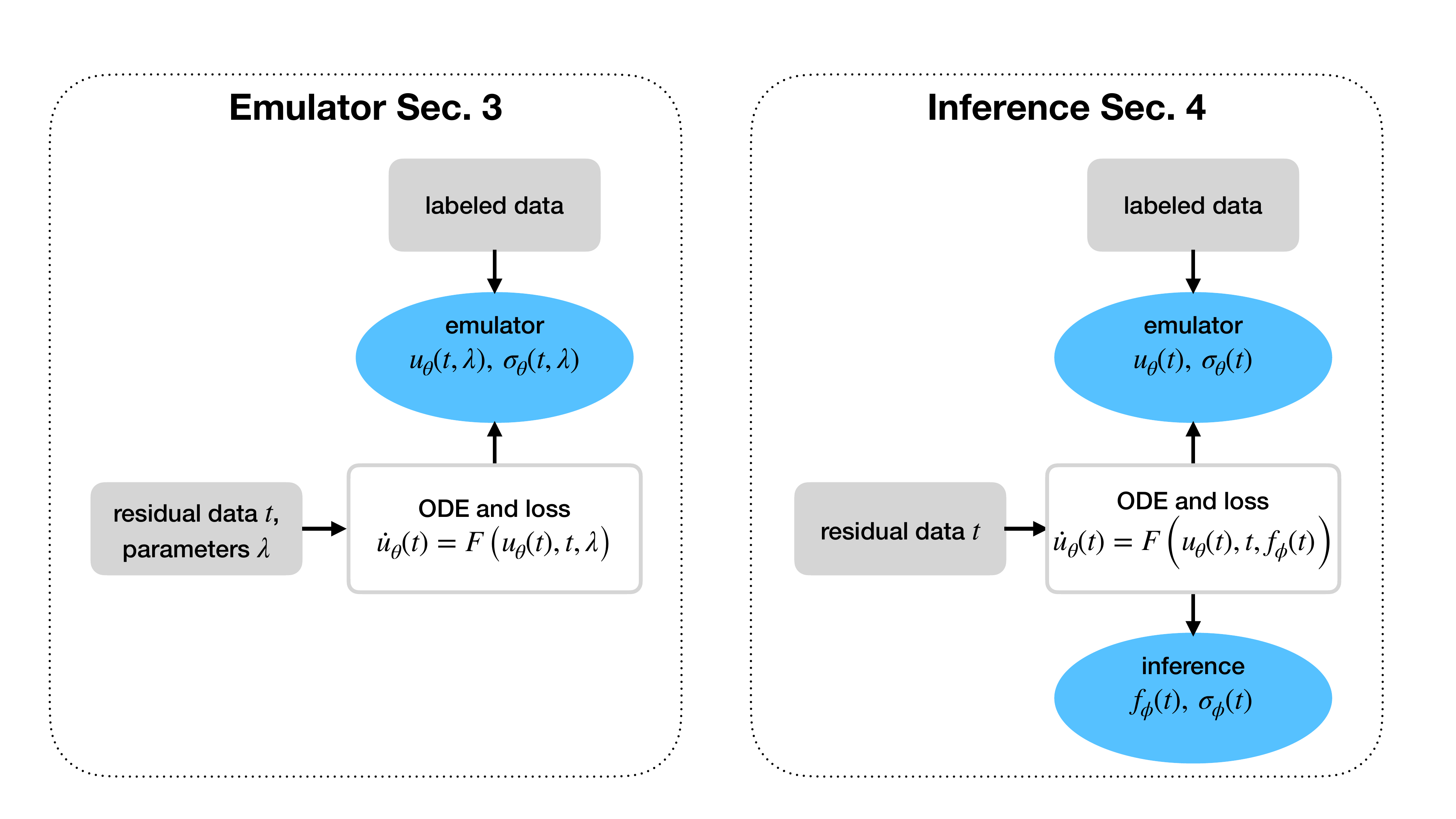}
    \caption{Illustration of the PINN emulation and inference setups.}
    \label{fig:schematic}
\end{figure}
%-----------------------------------------------

The previous section demonstrates that PINNs can learn and emulate luminosity 
distances arising for a given parameterized Hubble function as a solution to a 
differential equation. Inference inverts this process. Now the errors on a dataset 
need to be mapped onto a corresponding uncertainty on the inputs, either discrete 
parameters or a neural network-represented free Hubble function, as we 
will do next.

The formulation of a differential equation with a free function $f_\phi(t) \approx f(t)$ to be represented by a neural network, similar to~\cite{Shukla2020PhysicsInformedNN}, expands the structure of 
Eq.\eqref{eq:generalODE} to
\begin{align}
	\dot{{u}}(t) = F({u(t)},t, f(t))
	\quad\text{with}\quad
	{u}(0) = {u}_0  \; .
 \label{eq:generalODEInverse}
\end{align}
We extract information on the differential equation including $f(t)$ by training a network $u_\theta(t)$ on the labeled data. This network should fulfill the differential equation with the true function $f(t)$. This function is approximated with a second network $f_\phi(t)$. Given $N$ labeled data points $(t, u)_i$ and $M$ residual points $\tilde{t}_j$ the training uses the loss functions
\begin{align}
    \loss_\text{Data} &= \frac{1}{N} \sum_{i=1}^N \left[ u_\theta(t_i) - u_i\right]^2 \notag \\
    \loss_\text{ODE} &= \frac{1}{M}\sum_{j=1}^M \left[ \dot{{u}}_\theta(\tilde{t}_j) - F({u_\theta(\tilde{t}_j)},\tilde{t}_j, f_\phi(\tilde{t}_j))\right]^2.
\end{align}
The data loss plays the same role as $\loss_\text{IC}$ in Eq.\eqref{eq:generalLoss}.
This ensures $f_\phi(t) \approx f(t)$ for all times covered by
the residual points, as long as $u_\theta$ is sufficiently
accurate. The information on $f(t)$ is
first extracted from the data using the network approximating the
differential equation via $u_\theta$. In a second step, the differential
equation is used to infer the function itself. In all numerical
experiments the losses are combined by alternating between epochs
using only one of them. The
network structure and training are illustrated in
Fig.~\ref{fig:schematic}.

%-----------------------------------------------
\begin{figure}[t]
	\includegraphics[width = 0.495\textwidth]{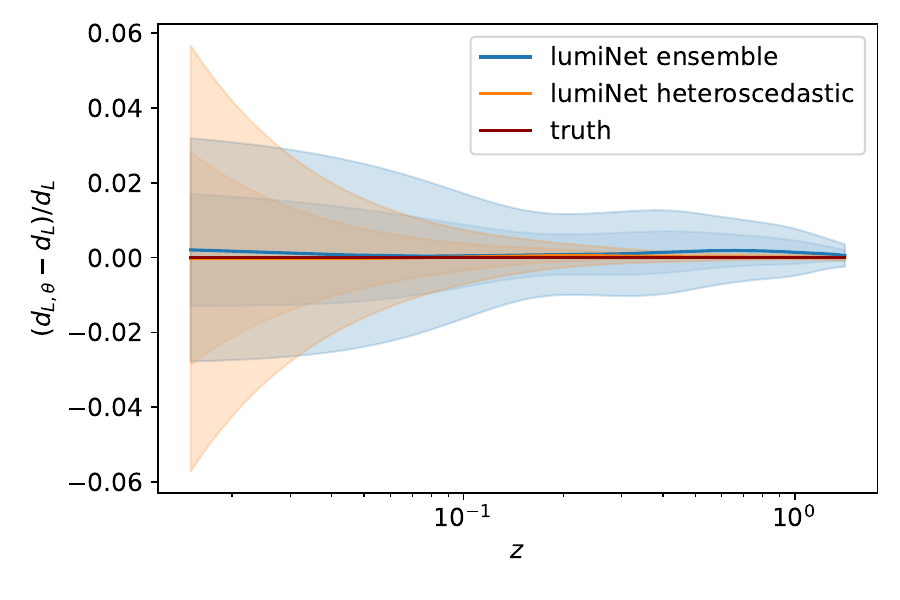}
	\includegraphics[width = 0.495\textwidth]{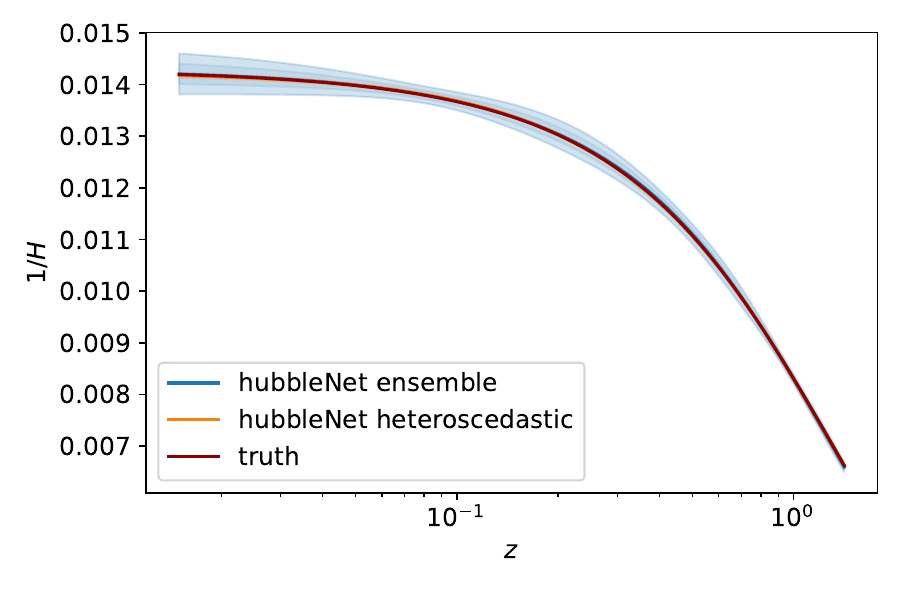}
	\caption{Reconstruction of the Hubble function using PINNs with an MSE loss.  The left plot depicts the luminosity distance approximation compared to the true value with an ensembling error bar derived from ten models. The right-hand side depicts the corresponding Hubble reconstruction.}\label{fig:inverseHubble_simData}
\end{figure} 
%-----------------------------------------------

Figure~\ref{fig:inverseHubble_simData} tests this setup for a cosmological model 
defined by Eq.\eqref{eq:hubbleParametrized} with $w$ fixed 
to the best-fit value of the Union2.1 dataset.
This reconstruction uses a dense network with five hidden layers with a 
width of $100$ nodes for both networks encoding the luminosity distance $d_\theta$ and 
the Hubble function, $H_\phi$. The reconstruction is performed using $10^4$ residual 
points and $10^3$ labeled data points, the same order of magnitude as current 
surveys~\cite{Scolnic_2022, Rubin:2023ovl}. For each epoch of training with the ODE 
loss ten epochs are trained with the data loss. Alternating between epochs of training 
on the ODE and training on data allows to control the relative weight between 
fulfilling the ODE and fitting data.
This reconstruction of the (inverse) Hubble function is performed without 
any input on the particular model used to generate the data. On this synthetic data set the Hubble function can be learned almost perfectly. 

%%%%%%%%%%%%%%%%%%%%%%%%%%%%%%%%%%%%%%%%%%%%%%%%
\subsection{Uncertainty estimation} 
\label{sec:PINNferenceUncert}

A key part of ML-inference is the control over uncertainties affecting the 
network training. As demonstrated in Sec.~\ref{sec:pinn_repuls}, a
repulsive ensemble of networks extracts a meaningful uncertainty, especially in 
regions with sparse data. To confirm this error estimate, we also 
use a heteroscedastic loss, in particular when there are no significant gaps in the 
data.

Combining the learned luminosity distance $\tilde{d}_{L,\theta}$, 
with uncertainty $\sigma_\theta$, and the ODE in  Eq.\eqref{eq:lumiODE},
every luminosity distance value contributes to the reconstruction of the Hubble function as
\begin{align}
	\frac{1+z_i}{\tilde{H}(z_i)} \approx \frac{\dd \tilde{d}_{L,\theta}(z_i)}{\dd z}- \frac{\tilde{d}_{L,\theta}(z_i)}{1+z_i}. \label{eq:supernovaHubbleFromLumi}
\end{align}
To include $\sigma_\theta$, both $\tilde{d}_{L,\theta}(z_i)$ and $\dd \tilde{d}_{L,\theta}(z_i)/\dd z$ need to be drawn from their respective probability 
distributions. By using a heteroscdastic loss the luminosity distance at each redshift 
is assumed to follow a normal distribution $\mathcal{N}(\tilde{d}_{L,\theta}(z_i), \sigma^2_\theta)$. 
Since samples of the luminosity distance are generated using a standard Gaussian,
the width of the 
derivative distribution is $\dd\sigma_\theta/\dd z$. Generating samples from these 
distributions and inserting them into Eq.\eqref{eq:supernovaHubbleFromLumi}, it is 
possible to generate a distribution of of Hubble function values for each redshift. 

A second network can then learn $\tilde{H}_\phi$ with an uncertainty $\sigma_\phi$ based 
on the luminosity distance network $d_{L,\theta}$, where both networks are trained to 
fulfill Eq.\eqref{eq:supernovaHubbleFromLumi} and fit the data. The uncertainty on 
$\tilde{H}_\phi$ is learned using the heteroscedastic loss of 
Eq.\eqref{eq:supernovaHeteroscedasticExplicit}. This uncertainty can be interpreted as 
the uncertainty on $(1+z)/\tilde{H}_\phi(z)$ under the assumption that for each 
redshift $d_{L,\theta}$ fulfills the differential equation correctly. This allows us 
to reduce the loss function to the expression 
\begin{align}
	\loss_\text{Hubble, het} = \frac{1}{N}\sum_{i=1}^{N} \left[\frac{\left( \dfrac{1+z_i}{\tilde{H}(z_i)} - \dfrac{1+z_i}{\tilde{H}_\phi(z_i)}\right)^2}{2\left(\sigma_\phi(z_i)\right)^2}+ \log\sigma_\phi(z_i)\right]. \label{eq:hubbleNetLossFunction}
\end{align}
The Hubble function is then approximated by a normal distribution in 
$(1+z_i)/\tilde{H}_\phi(z_i)$ with variance $\sigma^2_\phi(z_i)$. 

The combination of Eqs.\eqref{eq:supernovaHubbleFromLumi} 
and~\eqref{eq:hubbleNetLossFunction} allows us to optimize $H_\phi$ and 
$d_{L,\theta}$ simultaneously. The mean value and uncertainty of $d_{L,\theta}$ 
appear in the sampling of $d_L(z_i)$, allowing the network parameters $\theta$ 
to influence the loss. For every epoch trained using the differential equation loss, 
the PINN for the luminosity distance is also trained to match the labeled data. The 
ratio of labeled data epochs to ODE epochs is a training hyper-parameter.

%%%%%%%%%%%%%%%%%%%%%%%%%%%%%%%%%%%%%%%%%%%%%%%%
\subsection{Noisy data}
\label{sec:cosmodata}

To analyse real data we have to allow for noise in 
solving the inverse problem. We consider two datasets, 
Union2.1~\cite{suzuki_hubble_2012, amanullah_spectra_2010, kowalski_improved_2008} 
and Pantheon+~\cite{Scolnic_2022}. 
To use them as labeled training data, we convert them into luminosity distances with 
corresponding error bars using Eq.\eqref{eq:DistanceModulus}. Assuming that the data 
follows a multivariate normal distribution, %$\mathcal{N}({y},\Sigma)$, 
we can generate a set of luminosity distances per redshift using the mean and the covariance matrix from the actual data.

The resulting luminosity distances and the distribution of redshifts for the ensemble of synthetic datasets is depicted in Fig.~\ref{fig:supernovaData}. Typical errors 
are around 10\%, and the data becomes sparse towards large redshift. The newer
Pantheon+ dataset covers a larger range in redshifts and includes three times as many 
supernovae.

%-----------------------------------------------
\begin{figure}[b!]
	\includegraphics[width = 0.495\textwidth]{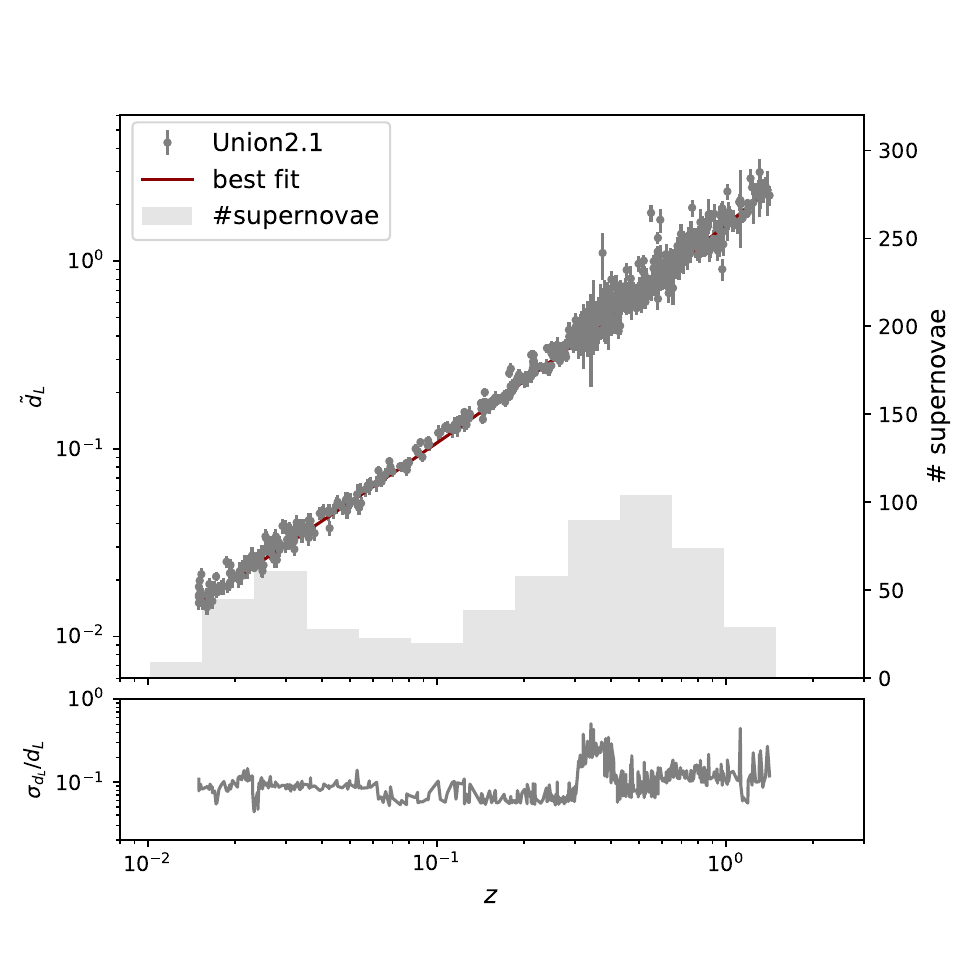}
	\includegraphics[width = 0.495\textwidth]{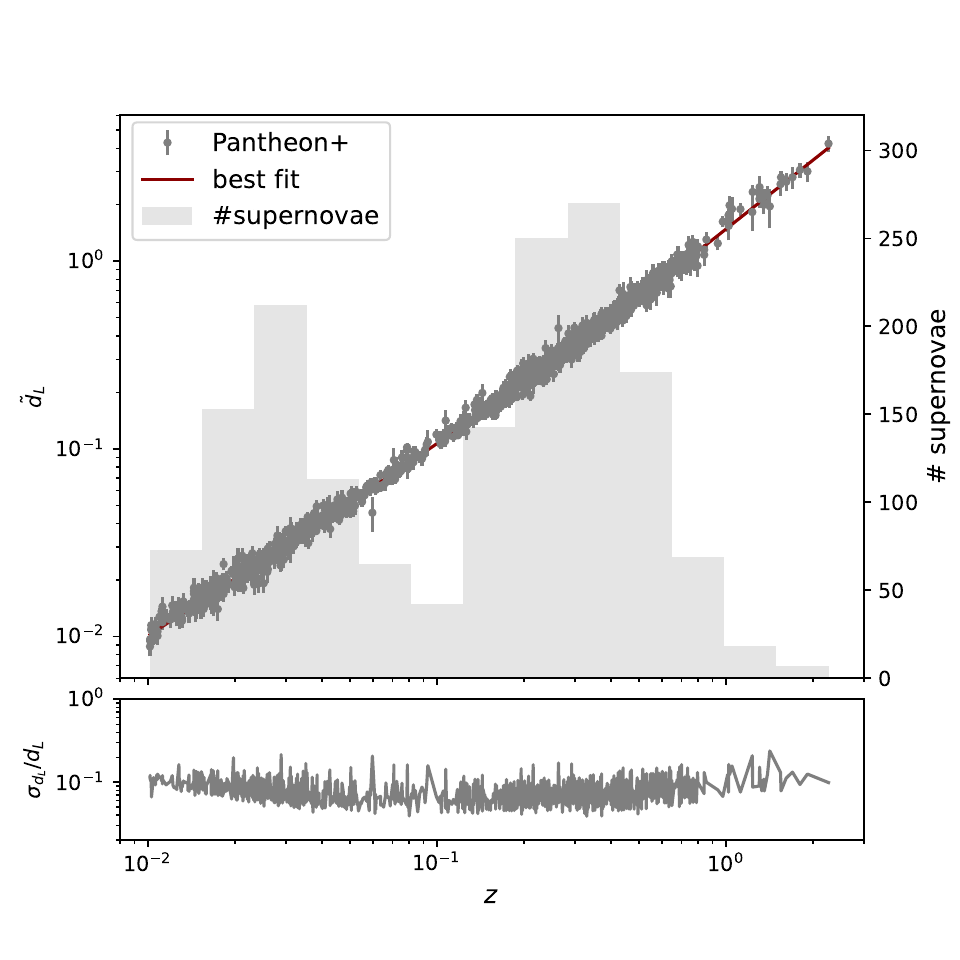}
	\caption{Generated redshift dependencies of the luminosity distance values of the Union2.1
    (left) and Pantheon+ data (right). The histograms capture the distribution of the supernovae in redshift. The lower sub-panels show the relative error bars on the luminosity distances. }\label{fig:supernovaData}
\end{figure} 
%-----------------------------------------------

In this section the luminosity distance is learned as $d_\theta$, using five 
layers with $100$ nodes each. This network is trained on the labeled data. The 
inverse Hubble function is modeled with a second network with five layers and 200 
nodes wide. As suggested in Ref.~\cite{Wang2023AnEG} we impose the boundary condition 
of the luminosity distance network by learning $(d_{L}/z)_\theta$ and multiplying by 
$z$ later. In addition, random Fourier features~\cite{Tancik2020FourierFL} 
significantly reduces the required training time. For each dataset, 
the training data for each epoch is generated 
from the luminosity distance distribution shown in Fig.~\ref{fig:supernovaData}. 
The resulting ensemble of luminosity distances scatters around the mean 
at each redshift, which can be captured by the heteroscedastic loss.

%-----------------------------------------------
\begin{figure}[t]
	\includegraphics[width = 0.495\textwidth]{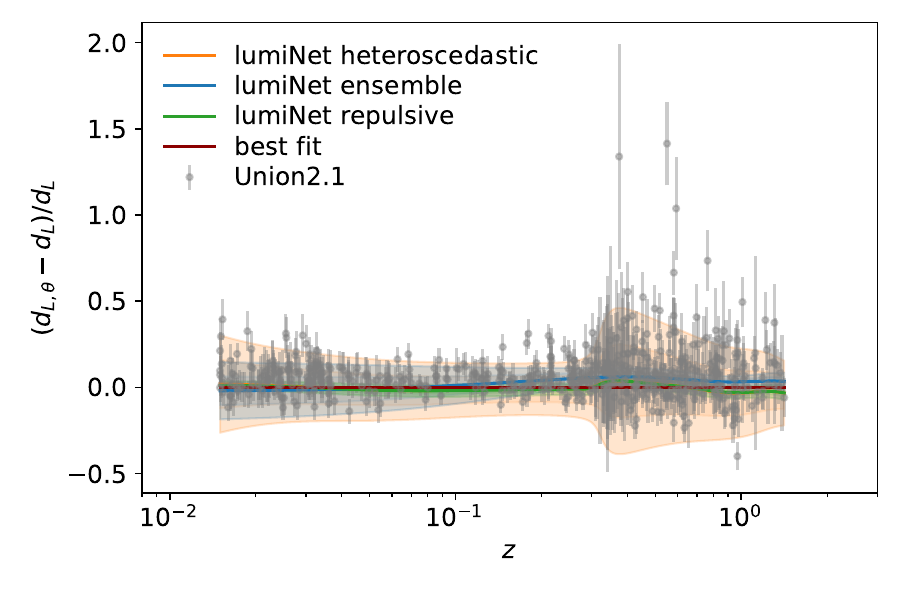}
    \includegraphics[width = 0.495\textwidth]{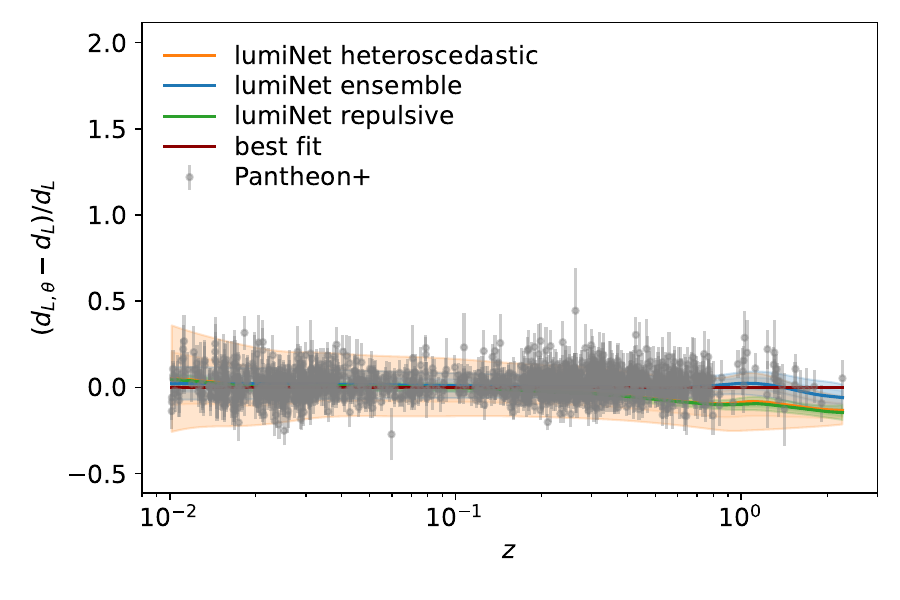}\\
	\includegraphics[width = 0.495\textwidth]{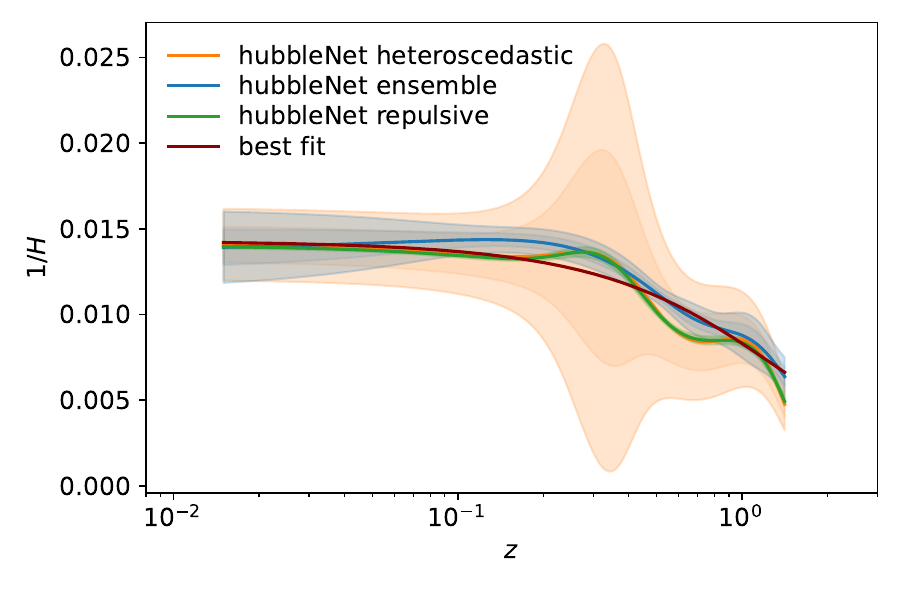}
    \includegraphics[width = 0.495\textwidth]{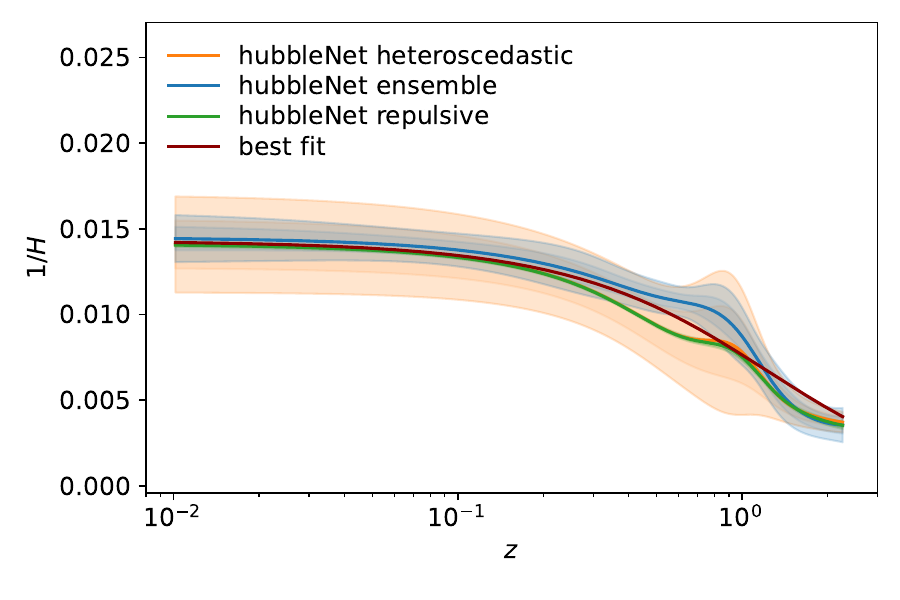}
	\caption{Top: PINN-learned luminosity distance from the labeled data, 
    derived from the Union2.1 (left) and Pantheon+ (right) data. Bottom: 
    learned inverse Hubble function from the two datasets.}
 \label{fig:inverseHubbleRealData}
\end{figure} 
%-----------------------------------------------

In Fig.~\ref{fig:inverseHubbleRealData} we show the reconstruction of the Hubble 
function from both datasets. We shown the learned luminosity distance and 
the the reconstructed Hubble function, comparing a heteroscedastic network, an ensemble of MSE networks 
and a repulsive ensemble. 
Similar to Sec.~\ref{sec:pinn_res}, the ensemble and the repulsive ensemble using the labeled data 
region does not capture the noise, whereas the heteroscedastic uncertainty of the 
luminosity distance does. The reconstructed Hubble function is consistent 
with a $w$CDM approximation of the Hubble function from a direct fit of a 
parameterized model. 

The sharp feature in the Hubble function reconstruction from the Union2.1 dataset can 
be understood from Eq.\eqref{eq:supernovaHubbleFromLumi}. The uncertainty of the  
Hubble function is approximately the quadratic mean of the uncertainty of the 
derivative of the luminosity distance and the uncertainty of the luminosity distance 
itself. Fast changes in the width and scatter of the labeled data points with 
redshift, see Fig.~\ref{fig:supernovaData}, leverage fast changes in the predicted 
error bars of the Hubble function. The sharp increase in the uncertainty of the 
reconstructed Hubble function at redshift $0.3$ corresponds to the change in the 
uncertainty in the luminosity distance leading to a maximum in the uncertainty. 

The reconstruction of the Hubble function in Eq.\eqref{eq:supernovaHubbleFromLumi} 
relies on the assumption that the network approximating the luminosity distances 
fulfills the differential equation exactly. The deviation from the a true solution can 
be approximated by inserting both networks into the differential equation. When 
applying PINN inference to our datasets, this deviation is small compared to the predicted uncertainties from the spread of the data.

%%%%%%%%%%%%%%%%%%%%%%%%%%%%%%%%%%%%%%%%%%%%%%%%
\subsection{Dark energy equation of state}

Finally, we convert the inferred, parameter-free Hubble function $H(a)/H_0$ to an 
equation of state function $w(a)$. Using the general relation~\cite{Takada:2003ef},
\begin{align}
\frac{H^2(a)}{H_0^2} = 
\frac{\Omega_m}{a^3} + (1-\Omega_{m}) \; 
\exp\left[ -3 \int_1^a da' \frac{1+w(a')}{a'}\right] \; ,
\end{align}
we extract $w(a)$ by differentiation,
\begin{align}
w(a) 
= -\frac{1}{3}\frac{\dd}{\dd\log a}\log \left[ \frac{H^2(a)}{H_0^2}-\frac{\Omega_m}{a^3}\right] - 1 \; .
\label{eqn_de_eos}
\end{align}
We use $\Omega_m = 0.28$, as suggested by the Union2.1 dataset. Naturally, this 
differentiation introduces a larger uncertainty when transitioning from the inferred 
Hubble function $H$ to the equation of state $w$.

%-----------------------------------------------
\begin{figure}[t]
	\includegraphics[width = 0.495\textwidth]{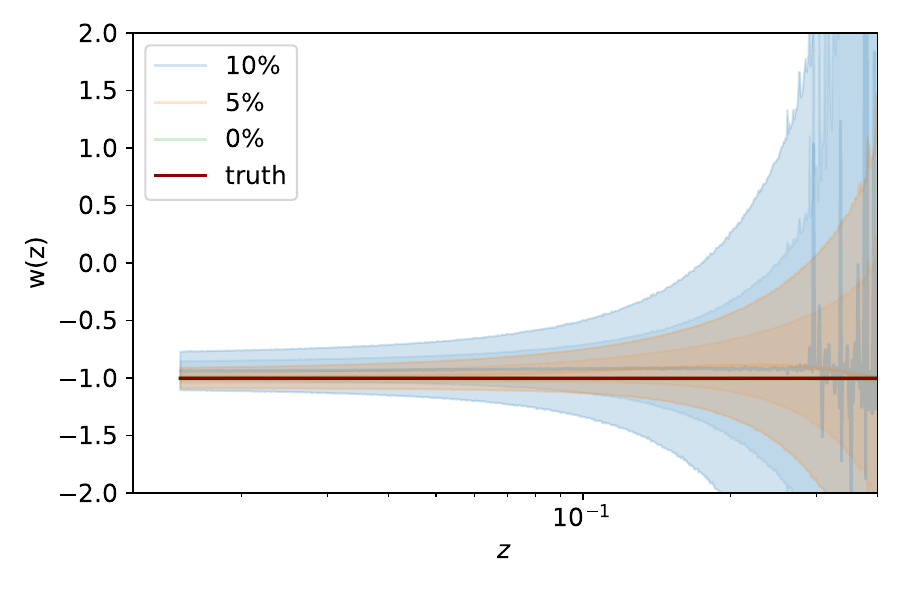}
    \includegraphics[width = 0.495\textwidth]{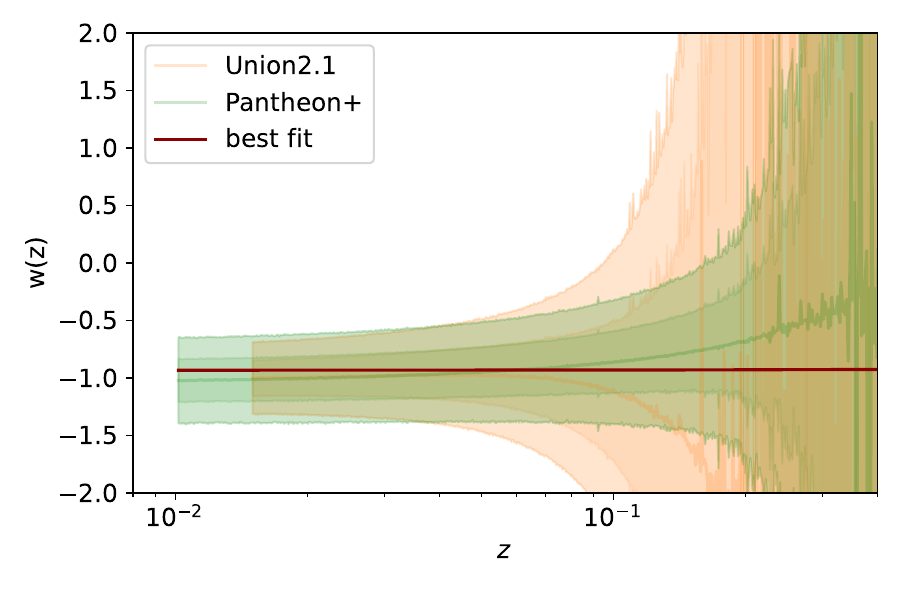}
	\caption{Inferred dark energy equation of state. The left panel uses simulated date with increasing assumed error bars. The right panel uses the Union2.1 and the Pantheon+ dataset, propagating the
 error bars estimated by the collaborations through the PINN-inference.}
 \label{fig:eosFromHubble}
\end{figure} 
%-----------------------------------------------

Performing a test on $10^3$ simulated  supernovae, uniformly in redshift, leads to the 
left panel of Fig.~\ref{fig:eosFromHubble}. It demonstrates that we can reconstruct 
$w(z)$ with small uncertainties. Increasing the observational uncertainty to $5$\% or 
$10$\%  shows a commensurate effect on uncertainty of the inferred $w(z)$. In terms of 
redshift, the uncertainty becomes large beyond $z\simeq0.3$ for realistic errors, 
which is partially caused by the uncertainty of the PINN far away from its 
initial conditions. But more importantly, dark energy has a small influence on the 
Hubble function at high redshift, rendering $w(z)$ effectively 
unconstrained. Technically, by approaching $H(a)^2 \simeq \Omega_m/a^3$ at 
sufficiently high redshifts leads to a diverging logarithmic derivative 
in Eq.\eqref{eqn_de_eos}.

In the right panel of Fig.~\ref{fig:eosFromHubble} we show the reconstruction of 
$w(z)$ from our two datasets.
The matter density for each of dataset is again assumed to be the best fit value. 
At small redshifts our inference method constrains $w(z)$ well, but 
the uncertainties of the labeled data do not leave any sensitivity 
beyond $z \gtrsim 0.3$.

%%%%%%%%%%%%%%%%%%%%%%%%%%%%%%%%%%%%%%%%%%%%%%%%
\section{PINNclusions}
\label{sect:summary}

Physics-informed neural networks are trained on the output of a parameterized system of differential equations. They can predict solutions for given parameters with a proper interpolation between parameter choices. This emulation of the space of ODE solutions 
provides tremendous speed-ups and therefore an excellent tool for statistical inference. The focus of our investigation was the error-awareness or uncertainty estimation of PINNs. For this purpose we have compared a heteroscedastic loss and repulsive ensembles, confirming that PINNs extrapolate into regions of sparse or low-quality data, while sensibly increasing their learned error in these regions. Testing these aspects with the harmonic oscillator as a toy
example confirms the fundamental behavior of PINNs.

The functionality of PINNs as emulators was then verified with luminosity distances as functions of redshift for a conventional dark-energy dominated Friedmann-Robertson-Walker universe. PINNs correctly predict the luminosity distance for a given redshift over a wide range of dark energy equation of state parameters, without solving a differential equation, or equivalently in this case, performing a numerical integration.

Using PINNs for inference rather than emulation requires a statistical inversion, i.e. a mapping of the experimental uncertainty back to the parameterization. Applied to the supernova example, PINNs allow for an uncertainty-aware reconstruction of the Hubble function without any predefined parameterization. The Hubble function is reconstructed by the PINN including an error estimate. They discover peculiarities in the data, such as the sudden increase in error in the Union-data set at $z\simeq0.3$, reflecting a large uncertainty in the reconstructed Hubble function. Re-expressing the Hubble-function with the dark energy equation of state function derived for a fixed matter density shows weaker constraints, as the increase in error is driven by the derivative transitioning from $H(a)$ to $w(a)$.

%%%%%%%%%%%%%%%%%%%%%%%%%%%%%%%%%%%%%%%%%%%%%%%%
\section*{Acknowledgements}

We would like to thank Manuel Haussmann for recommending repulsive ensembles to us, Theo Heimel for his advice and his implementation of the repulsive ensmbles, Ullrich K\"othe for valuable technical help with PINNs, and Benedikt Schosser for help with the Pantheon+ dataset.
This work was supported by the Deutsche Forschungsgemeinschaft (DFG, German Research Foundation) under
Germany's Excellence Strategy EXC 2181/1 - 390900948 (the Heidelberg STRUCTURES Excellence Cluster). 

\bibliography{tilman,references}
\end{document}